\documentclass[preprint]{elsarticle}

\journal{Marine Structures}

\usepackage{subcaption}
\usepackage{amsmath,amsfonts,amssymb}
\usepackage{mathtools}
\usepackage{leftidx}
\usepackage[group-minimum-digits = 6]{siunitx}
\usepackage{pstool}
\usepackage{import}
\usepackage{multicol, multirow}
\usepackage{xspace}
\usepackage{url}
\usepackage[pdfpagelabels]{hyperref}
\hypersetup{
   pdftitle={Semi-Submersible Wind Turbine Hull Shape Design for a Favorable System Response Behavior},    
   pdfauthor={Frank Lemmer},
   pdfcreator={PDFLaTeX},
   colorlinks=false,
   linkcolor=black,
   urlcolor=black,
   citecolor=black,
   pdfborder=0 0 0,
   plainpages=false}

\usepackage{eurosym}
\DeclareSIUnit{\EUR}{\text{\euro}}

\newcommand{\vek}[1]{\mbox{\boldmath${#1}$}} 
\newcommand{\mat}[1]{\mbox{\boldmath${#1}$}} 
\newcommand{\vol}{\triangledown}

\newcommand{\fowt}{\gls{fowt}\xspace}
\newcommand{\fowts}{\glspl{fowt}\xspace}

\newcommand{\dofs}{\glspl{dof}\xspace}
\newcommand{\rwt}{DTU~\SI{10}{MW}~\gls{rwt}\xspace}

\newcommand{\slow}{\gls{slow}\xspace}

\renewcommand{\arraystretch}{1.2} 

\newcommand{\clrI}{blue\xspace} 
\newcommand{\clrII}{red\xspace}
\newcommand{\clrIII}{green\xspace}

\renewcommand{\arraystretch}{1.2} 

\usepackage[acronym,toc,nonumberlist]{glossaries} 
\glssetwidest[0]{} 
\glssetwidest[1]{1234567890} 

\makenoidxglossaries

\newacronym			{lidar}		{LiDAR} 	{Light Detection And Ranging}
\newacronym			{fowt}		{FOWT} 		{Floating Offshore Wind Turbine}
\newacronym			{iea}		{IEA}		{International Energy Agency}
\newacronym			{iec}		{IEC}		{International Electrotechnical Commission}
\newacronym			{swl}		{SWL}		{Still Water Level}
\newacronym			{trl}		{TRL}		{Technology Readiness Level}
\newacronym			{tlp}		{TLP}		{Tension Leg Platform}
\newacronym[plural=DoFs,firstplural=Degrees of Freedom~(DoFs)]			{dof}		{DoF}		{Degree of Freedom}
\newacronym			{rwt}		{RWT}		{Reference Wind Turbine}
\newacronym			{nrel}		{NREL}		{National Renewable Energy Laboratory, Boulder, USA}
\newacronym			{ntnu}		{NTNU}		{Norwegian University of Science and Technology}
\newacronym			{nasa}		{NASA}		{National Aeronautics and Space Administration, USA}
\newacronym			{ntua}		{NTUA}		{National Technical University of Athens}
\newacronym			{mit}		{MIT}		{Massachussetts Institute of Technology}
\newacronym			{dtu}		{DTU}		{Technical University of Denmark}
\newacronym			{dhi}		{DHI}		{Danish Hydraulic Institute}
\newacronym			{ecn}		{ECN}		{Energy Research Center of the Netherlands}
\newacronym			{ecnnantes} {ECN}		{Ecole Centrale de Nantes, France}
\newacronym			{marin}		{MARIN}		{Marine Research Institute Netherlands}
\newacronym			{swe}		{SWE}		{Stuttgart Wind Energy}
\newacronym			{eawe}		{EAWE}		{European Academy of Wind Energy}
\newacronym			{lheea}		{LHEEA}		{Research Laboratory in Hydrodynamics, Energetics \& Atmospheric Environment, Nantes, France}
\newacronym			{cener}		{CENER}		{National Renewable Energy Centre of Spain}
\newacronym			{ife}		{IFE}		{Institute for Energy Technology, Norway}
\newacronym			{lcoe}		{LCOE}		{Levelized Cost of Energy}
\newacronym			{oop}		{OoP}		{Out-of-Plane}
\newacronym			{ip}		{IP}		{In-Plane}
\newacronym			{psd}		{PSD}		{Power Spectral Density}
\newacronym			{1p}		{1p}		{Once-Per-Revolution}
\newacronym			{3p}		{3p}		{Three-Times-Per-Revolution}
\newacronym			{oc3}		{OC3}		{Offshore Code Comparison Collaboration}
\newacronym			{oc4}		{OC4}		{Offshore Code Comparison Collaboration, Continued}
\newacronym			{oc5}		{OC5}		{Offshore Code Comparison Continuation, Continued, with Correlation}
\newacronym			{mbs}		{MBS}		{Multibody System}
\newacronym			{fe}		{FE}		{Finite Element}
\newacronym			{bem}		{BEM}		{Blade Element Momentum}
\newacronym			{tsr}		{TSR}		{Tip Speed Ratio}
\newacronym			{cfd}		{CFD}		{Computational Fluid Dynamics}
\newacronym			{gdw}		{GDW}		{Generalized Dynamic Wake}
\newacronym			{lti}		{LTI}		{Linear Time-Invariant}
\newacronym			{rao}		{RAO}		{Response Amplitude Operator}
\newacronym	[plural=EQM,firstplural=Equations of Motion~(EQM)]		{eqm}		{EQM}		{Equation of Motion}
\newacronym			{dft}		{DFT}		{Discrete Fourier Transform}
\newacronym			{idft}		{IDFT}		{Inverse Discrete Fourier Transform}
\newacronym			{siso}		{SISO}		{Single-Input-Single-Output}
\newacronym			{mimo}		{MIMO}		{Multi-Input-Multi-Output}
\newacronym			{ode}		{ODE}		{Ordinary Differential Equation}
\newacronym			{fls}		{FLS}		{Fatigue Limit State}
\newacronym			{uls}		{ULS}		{Ultimate Limit State}
\newacronym			{dlc}		{DLC}		{Design Load Case}
\newacronym			{lc}		{LC}		{Load Case}
\newacronym			{ntm}		{NTM}		{Normal Turbulence Model}
\newacronym			{pdf}		{PDF}		{Probability Density Function}
\newacronym			{cdf}		{CDF}		{Cumulated Distribution Function}
\newacronym			{std}		{STD}		{Standard Deviation}
\newacronym			{rhpz}		{RHPZ}		{Right Half-Plane Zero}
\newacronym			{ipc}		{IPC}		{Individual Pitch Control}
\newacronym			{cpc}		{CPC}		{Collective Pitch Control}
\newacronym			{mpc}		{MPC}		{Linear Model-Predictive Control}
\newacronym			{nmpc}		{NMPC}		{Nonlinear Model-Predictive Control}
\newacronym			{lqr}		{LQR}		{Linear Quadratic Regulator}
\newacronym			{dac}		{DAC}		{Disturbance Accomodating Controller}
\newacronym			{tmd}		{TMD}		{Tuned Mass Damper}
\newacronym			{pi}		{PI}		{Proportional-Integral}
\newacronym			{hawt}		{HAWT}		{Horizontal-Axis Wind Turbine}
\newacronym			{vawt}		{VAWT}		{Vertical-Axis Wind Turbine}
\newacronym			{dae}		{DAE}		{Differential Algebraic Equation}
\newacronym			{pde}		{PDE}		{Partial Differential Equation}
\newacronym			{rna}		{RNA}		{Rotor-Nacelle Assembly}
\newacronym			{ol}		{OL}		{Open Loop}
\newacronym			{cl}		{CL}		{Closed Loop}
\newacronym			{mor}		{MOR}		{Model Order Reduction}
\newacronym			{sid}		{SID}		{Standard Input Data}
\newacronym			{zoh}		{ZOH}		{Zero-Order Hold}
\newacronym			{cm}		{CM}		{Center of Mass}
\newacronym			{cb}		{CB}		{Center of Buoyancy}
\newacronym			{cf}		{CF}		{Center of Flotation}
\newacronym			{rms}		{RMS}		{Root Mean Square}
\newacronym			{afosp}		{AFOSP}		{Alternative Floating Platform Designs for Offshore Wind Turbines using Low Cost Materials}
\newacronym			{slow}		{SLOW}		{Simplified Low-Order Wind Turbine}
\newacronym			{rga}		{RGA}		{Relative Gain Array}
\newacronym			{gm}		{GM}		{Gain Margin}
\newacronym			{pm}		{PM}		{Phase Margin}
\newacronym			{cad}		{CAD}		{Computer-Aided Design}
\newacronym			{qtf}		{QTF}		{Quadratic Transfer Function}
\newacronym			{TF}		{TF}		{Transfer Function}
\newacronym			{svd}		{SVD}		{Singular Value Decomposition}
\newacronym			{del}		{DEL}		{Damage-Equivalent Load}
\newacronym			{dnvgl}		{DNV-GL}	{Det Norske Veritas - Germanischer Lloyd}
\newacronym			{pso}		{PSO}		{Particle-Swarm Optimization}
\newacronym			{ga}		{GA}		{Genetic Algorithm}
\newacronym			{sqp}		{SQP}		{Sequential Quadratic Programming}
\newacronym			{mdo}		{MDO}		{Multidisciplinary Design Optimization}
\newacronym			{eog}		{EOG}		{Extreme Operating Gust}
\newacronym			{dp}		{DP}		{Dynamic Positioning}
\newacronym			{hil}		{HIL}		{Hardware-in-the-Loop}
\newacronym			{asme}		{ASME}		{American Society of Mechanical Engineers}

\glsaddall

\begin{document}

\begin{frontmatter}

\title{Semi-Submersible Wind Turbine Hull Shape Design for a Favorable System Response Behavior}

\author{Frank Lemmer} 
\ead{lemmer@ifb.uni-stuttgart.de}
\author{Wei Yu}
\author{Kolja Müller\fnref{1}}\fntext[1]{Independent researcher}
\author{Po Wen Cheng}
\address{University of Stuttgart (SWE), Allmandring 5B, 70569 Stuttgart, Germany}


%

\begin{abstract}
Floating offshore wind turbines are a novel technology, which has reached, with the first wind farm in operation, an advanced state of development. The question of how floating wind systems can be optimized to operate smoothly in harsh wind and wave conditions is the subject of the present work. An integrated optimization was conducted, where the hull shape of a semi-submersible, as well as the wind turbine controller were varied with the goal of finding a cost-efficient design, which does not respond to wind and wave excitations, resulting in small structural fatigue and extreme loads. 

The optimum design was found to have a remarkably low tower-base fatigue load response and small rotor fore-aft amplitudes. Further investigations showed that the reason for the good dynamic behavior is a particularly favorable response to first-order wave loads: The floating wind turbine rotates in pitch-direction about a point close to the rotor hub and the rotor fore-aft motion is almost unaffected by the wave excitation. As a result, the power production and the blade loads are not influenced by the waves. A comparable effect was so far known for Tension Leg Platforms but not for semi-submersible wind turbines. 
The methodology builds on a low-order simulation model, coupled to a parametric panel code model, a detailed viscous drag model and an individually tuned blade pitch controller. The results are confirmed by the higher-fidelity model FAST. A new indicator to express the optimal behavior through a single design criterion has been developed.
\end{abstract}


\end{frontmatter}


\section{Introduction}\label{sec:intro}
First concept and feasibility studies for \fowts appeared more than 15 years ago~\cite{Henderson2000}. Since then, prototypes of spars, semi-submersibles, barges and \glspl{tlp} have been built~\cite{Tande2015}. The present paper aims at particular design indicators for semi-submersibles for a favorable response behavior in wind and waves. The next sections will first provide an introduction of the inherent \fowt dynamic characteristics followed by a review of published design procedures.

\subsection{System dynamics of floating wind turbines}\label{sec:dynamics}
The commonly employed dynamic aero-hydro-elastic models for \fowts involve assumptions comparable to the FAST model by \gls{nrel}~\cite{FAST8}. This open-source tool is used as a reference model in the present work. 

Most models employ elastic \glspl{mbs} of reduced order through an approximation of the tower and blade deformation with a superposition of their respective mode shapes. FAST has in this work 25 enabled \dofs. The low-order model, on the other hand, has six \dofs, covering only the 2D planar motion of the system. The term ``low-order'' refers in this work to the number of \dofs of the dynamic equations of motion. Coupled dynamic models consider, next to the elastic forces, centrifugal, gyroscopic and Coriolis forces, which are significant for systems of large rigid-body motions. The coupling effects are important for the system dynamics, because the elastic body mode shapes vary when coupled into a multibody system. An example is the coupling of the tower flexibility with the floater rigid body modes or the controller closed-loop dynamics with the fore-aft mode. 

The wind turbine blade pitch controller is especially critical for \fowts: A standard rotor-speed controller for above-rated wind speeds will pitch the blades when the rotor speed exceeds its rated value. In the case of~\fowts, this feedback loop can imply, as a side-effect, that the tower or the platform experiences large excursions, due to the aerodynamic properties of the rotor. When the relative wind speed~(the one seen by the rotor) increases, the controller will pitch the blades towards feather~(increasing blade pitch angle) and thereby reduce the aerodynamic rotor torque. As a consequence, the thrust also decreases. This means, on the other hand, that an oscillation of the platform in pitch will become unstable if the controller reacts sufficiently fast to the sinusoidally oscillating relative wind speed, see~\cite{Jonkman2008,Larsen2007}.

External excitations arise from wind and waves. The wind turbulence excites the low-frequency platform modes as well as the blades and the tower-top at multiples of the rotor rotational frequencies. Dynamic effects due to the fore-aft motion of the rotor, accelerating the bulk flow across the turbine are subject to current research, i.e.~\cite{Hansen2014,Tran2015,Lennie2017}. Hydrodynamic forces, important for the system dynamics of semi-submersibles, are the Froude-Krylov wave loads including diffraction, viscous drag excitation, viscous drag damping~(both from Morison's drag term)~\cite{Matha2019a,Massie2001,mccormick2010} and second-order slow-drift forces at the difference-frequency of bichromatic waves from potential flow theory, i.e.~\cite{Faltinsen1993,Duarte2014}. The first-order wave force coefficient can feature an attenuation range for semi-submersibles, also called the ``wave cancellation effect'', see~\cite{Hanna1986} and \cite[p.~290]{Patel1989}. For a representation of the overall \fowt response, the mooring line dynamics might be of negligible influence in the case of catenary mooring lines~\cite{Azcona2016a,Matha2016a}.

\subsection{Floating wind turbine design}
Several schemes for the sequential design of \fowts have been published in~\cite{Huijs2013,Azcona2013,Fernandez,Crozier2011,Liu2018}. The report of the European project LIFES50+\footnote{\url{http://www.lifes50plus.eu}, accessed on July 18, 2018.}~\cite{Muller2015} provides a comprehensive literature review of the \fowt design process and summarizes the findings in a consolidated scheme consisting of three stages. This work does not introduce a new design procedure, but a favorable effect that becomes visible in an integrated simulation study, covering a range of system parameters. In order to understand the existing constraints and barriers for an integrated design procedure, these are outlined in this section, before addressing the present design and simulation methodology.

Commonly, the design of a floating platform starts with static calculations, the so-called spreadsheet design phase. In this phase, the main dimensions of the structure can be determined, satisfying the requirement of hydrostatic restoring with an approximate structural dimensioning. It includes also frequency-domain analyses of the floater rigid body and the mooring lines. The first stage terminates with first experiments and an approximation of the expected cost of energy. In the subsequent stage, the first coupled dynamic simulations take place, which require a wind turbine model and a controller. 

The manifold of design constraints of a \fowt are represented in various standards and guidelines~\cite{Atcheson2016}. Published design studies have addressed the structural integrity, dynamic behavior, manufacturing, installation and marine operations, among others. An academic concept design study across different platform types was presented in~\cite{Lefebvre2012}, outlining the main assumptions and design constraints. A spar platform out of concrete was designed, simulated and tested in~\cite{Molins2014a,Matha2015}. The constraints and options for manufacturing of that design are discussed and summarized in~\cite{Molins2018}. A comparable design of reduced draft was presented in~\cite{DeGuzman2018} with a focus on seakeeping of the floater and marine operations. The peculiar loads of a steel semi-submersible with submerged counter-weight are calculated in~\cite{Pereyra2018} with a focus on the cables, sustaining the counter-weight. A semi-submersible for a very large wind turbine was recently presented in~\cite{Liu2018}.

Further studies focused on methods for obtaining the structural stresses of the floater in extreme and fatigue load cases. One of the first approaches was presented for a steel semi-submersible in~\cite{Aubault2009}. The difficulty of obtaining the structural loads is due to the fact that most available coupled simulation codes assume the floater is rigid, which makes it impossible to calculate internal stresses. Ways to overcome this limitation were presented in~\cite{Luan2013,Luan2017,Borg2017,Nygaard2015}. These methods involve either additional \dofs of the dynamic model or additional engineering steps in decoupled analyses. Often, the tower-base bending moment, which is available in standard simulation models, is taken as representative structural load in optimization studies. It is influenced by both, wind and wave loads.

First integrated optimization studies for \glspl{tlp} and spars were presented in~\cite{Sclavounos2007}. A coupled frequency-domain model was used to calculate the tower-top acceleration for various shapes of the floating platform and different slack and taught mooring configurations. Later, optimization studies on spar platforms were presented~\cite{Fylling2011}, including load calculations and a cost model. An adaptation of the wind turbine controller was first included using the same simulation model as in the present work in~\cite{Sandner2014,Lemmer2017}.

A large optimization study across different platform types was published in~\cite{Hall2013} and later extended in~\cite{Karimi2018}. It includes the hull shape and mooring line design and a genetic optimizer. In that work, a frequency-domain model is derived from the code FAST~\cite{FAST8}, with a linear representation of the hydrodynamic viscous damping but without representing the wind turbine controller. The objective was also here a low nacelle acceleration, next to the estimated cost. 
The objective of the present work, as well as the design and simulation methodology will be outlined in the next section.

\subsection{Parametric design and simulation approach}
The methodology of the present work builds on parametric design routines~(i.e. estimating steel thicknesses for structural integrity) on the one hand and parametric simulation models on the other hand. The design task in this paper considers a variation of the hull shape of a three-column concrete semi-submersible with varying draft and the wind turbine controller.

\begin{figure}[htb]
\includegraphics[scale=1]{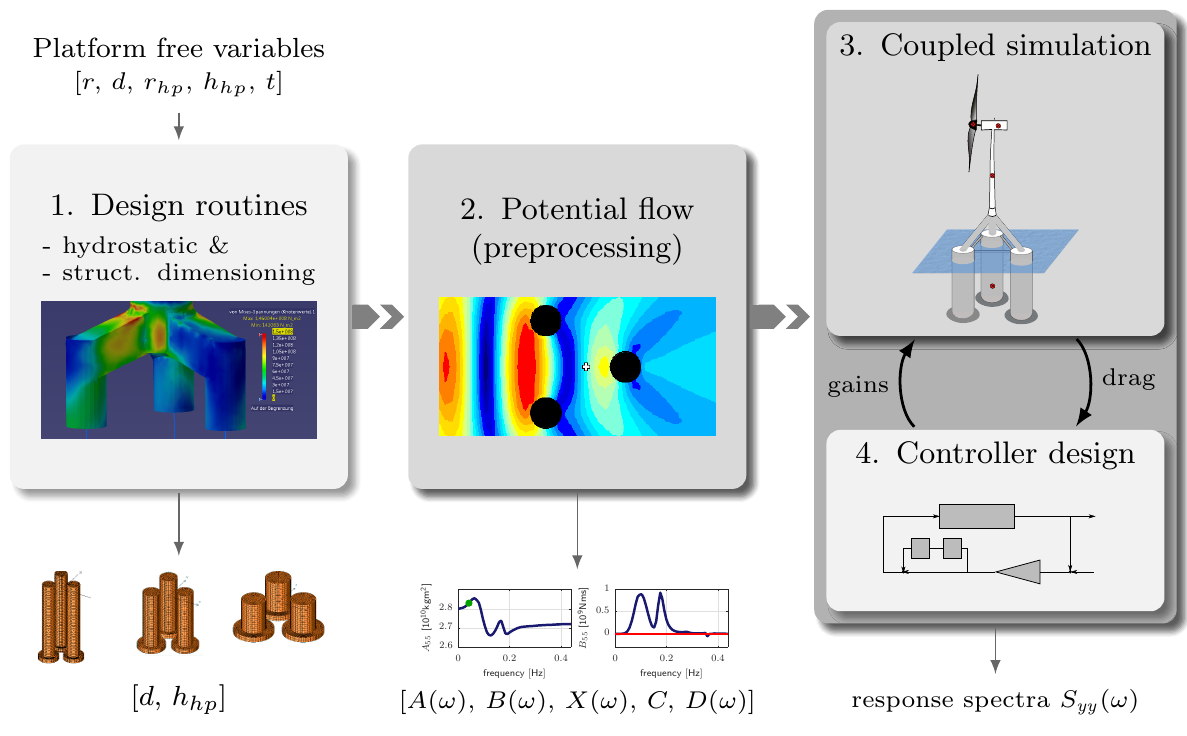}\\
\captionof{figure}{Parametric~\gls{fowt} system design and brute force optimization scheme.}\label{fig:FlowChart}
\end{figure}

Figure~\ref{fig:FlowChart} shows the main steps: The initial free variables parameterize the hull shape. As a function of these, a reasonable design space will be defined, with only few free variables to clearly visualize, interpret and understand the results, as shown in the next section. Based on this initial design space, the spreadsheet design routines ensure that each design satisfies basic constraints, related to the structural dimensions and the hydrostatic restoring, which is subject of Section~\ref{sec:parametricdesign}. The result of this step is a design space of only two dimensions. The designs range from a slender, deep-drafted semi-submersible to a design of large breadth, large column radius and shallow draft. All designs are three-column semi-submersibles with heave plates. Hence, no discontinuities are present. 

An adapted low-order aero-hydro-servo-elastic simulation model will be used for the analysis, as introduced in Section~\ref{sec:slow}. It includes all effects relevant for the main system dynamics, which are sought to be optimized. Therefore, effects like the floater structural elasticity or advanced aerodynamic effects are not considered. For all shapes, the first-order hydrodynamic coefficients are calculated through a parameterized panel code model. These coefficients are input to the low-order model. The model includes Newman's approximation to account for second-order slow-drift forcing. It includes also member-based viscous drag coefficients. The model is described in detail in~\cite{Lemmer2018e,Lemmer2018a}. 

The viscous heave plate drag model is rather detailed because it influences the fore-aft motion and the coupling with the controller. The vertical heave plate drag coefficients are parameterized as a function of the Keulegan-Carpenter number~$\mathit{KC}=vT/D$, which depends on the significant member length~$D$, the response velocity magnitude~$v$ and its period~$T$. 

The wind turbine blade-pitch controller design routine is a function of a linear state-space model, including the viscous drag coefficients. For this reason, the system response has to be solved for iteratively because the controller depends on the drag coefficients, which in turn depend on the response magnitude. Examples for the analyzed results of Section~\ref{sec:results} are the fatigue and extreme loads under realistic loading conditions, the modal properties and the harmonic response behavior to wind and waves.

The results of Section~\ref{sec:analysis} will analyze and visualize the distinct dynamic properties across the design space. This analysis helps to understand the effects, which drive load and power fluctuations. From the findings, a new design objective is derived, extending the one of the nacelle acceleration used in previous works. The objective is focused on the dynamic system response while the component design is out of the scope of this paper.

\section{Design Space}\label{sec:designspace}
\begin{figure}[htb]
 \centering
 \small
    \def\svgwidth{0.6\textwidth}
    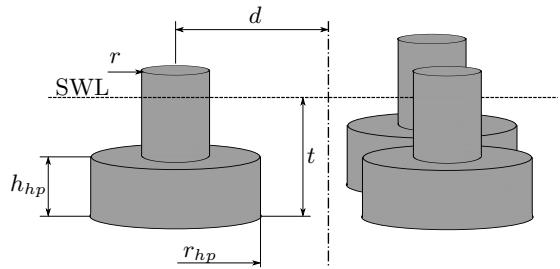
 	\caption{Free variables for parametric hull shape design,~\cite{Lemmer2018a}.}%
 	\label{fig:DesignSketch}%
\end{figure}

The selected hull shape parameters include the column spacing from the platform centerline~$d$, the column radius~$r$, the heave plate height~$h_{hp}$, the ratio of heave plate radius to column radius~$\hat{r}_{hp}=r_{hp}/r$ and the draft~$t$, see Figure~\ref{fig:DesignSketch}.  

The column radius~$r$ is defined to be \SI{52}{\percent} of the maximum possible column radius~$\hat{r}_\mathit{max}{=}d\sqrt{3}/2$ at which the three columns would touch each other. The heave plate-to-column radius ratio is intended to be~$\hat{r}_\mathit{hp,in}{=}2$ for all designs in order to ensure that the heave plates are large enough to effectively alter vertical added mass and Froude-Krylov forces. The heave plate-to-column radius ratio is altered, however, for the designs where the heave plates of the different columns would get too close to each other through an exponential function 
\begin{equation}
\hat{r}_\mathit{hp}=r_\mathit{hp}/r=(r_\mathit{max}(d)-1)(1-\text{exp}\left(\frac{\hat{r}_\mathit{hp,in}-1}{\hat{r}_\mathit{max}-1}\right) + 1.
\end{equation}
This ensures that no designs with unrealistically large heave plates result, which might not be manufacturable. The distance between overall center of mass and metacenter~$\mathit{GM}$, i.e.~\cite{mccormick2010}, determines the restoring as
\begin{equation}
C_{55}=\rho g \vol \mathit{GM},
\end{equation}
with the displaced volume~$\vol$ of density~$\rho$ and the gravity constant~$g$. Ballasting is such that zero trim is achieved. Thus, for given column diameters, a required~$C_{55}^*$ can be obtained by increasing the draft~$t$, which lowers the center of gravity more than it raises the center of buoyancy and therefore increases~$C_{55}$. 

An upper draft limit of~\SI{80}{m} was considered in the a-priori definition of the bounds of the free variables. Hence, any shape considered in this study can be described by the two free variables column spacing~$d$ and heave plate height~$h_{hp}$. This means that the design space (of the free variables) is Cartesian and the range of every variable does not depend on the values of the others. The small design space of feasible platforms makes a visualization and thorough analysis of the results possible. Table~\ref{tab:parameters} lists the free variables and the dependent variables.

\renewcommand{\arraystretch}{0.8}
\begin{table}[htp]
\centering
\caption{\fowt hull shape design parameters~\cite{Lemmer2018a}.}\label{tab:parameters}
\vspace{0.5em}
\begin{tabular}{ll}
\hline
Free variables & Dependent variables\\
\hline
\textbullet\: Column spacing~$d$	& \textbullet\:  Column radius~$r$ 	\\
\textbullet\: Heave plate height~$h_{hp}$  	& \textbullet\:  Heave plate radius~$r_{hp}$ 	\\
	& \textbullet\:  Draft~$t$ 	\\
  	& \textbullet\:  Steel tripod strut width~\& \\
  	& \hspace{1.11em}sheet thickness 	\\
	& \textbullet\:  Ballast mass 	\\
	& \textbullet\:  Platform mass distribution 	\\
	& \textbullet\:  Mooring line fairleads position	\\
	& \textbullet\:  Wind turbine controller gains\\
\hline
\end{tabular}
\end{table}

The semi-submersible main dimensions of the~2D design space are shown in Figure~\ref{fig:DesignSpace} with a column spacing range~$d=15.0(1.0)\SI{24.0}{m}$ and a heave plate height range~$h_{hp}=1.0(3.5)\SI{8.0}{m}$. The designs range from slender, rather ballast-stabilized towards buoyancy-stabilized shallow-drafted semi-submersibles. Larger column spacings than the ones considered are expected to result in excessive bending stresses in the tripod structure. The variable heave plate height has the main effect of modifying the vertical Froude-Krylov forcing. Thus, this variable may be able to facilitate the above-mentioned wave cancellation effect. The images on top of Figure~\ref{fig:DesignSpace} show that the column radius is largest for the lowest draft. The cost increases generally for increasing column radii~$r$ but decreases for the buoyancy-stabilized ones of shallow draft. The assumptions for the material cost estimation will be given in Section~\ref{sec:parametricdesign}, it is roughly proportional to the submerged volume. The three designs shown in Figure~\ref{fig:DesignSpace} will be selected in a number of the upcoming analyses and indicated by~``deep draft'', ``medium draft'', and ``shallow draft''. 

\setlength{\unitlength}{\textwidth}%
\begin{figure}[htb]
\centering
\begin{picture}(1.0, 0.7)
\put(0,0){
\includegraphics[scale=1]{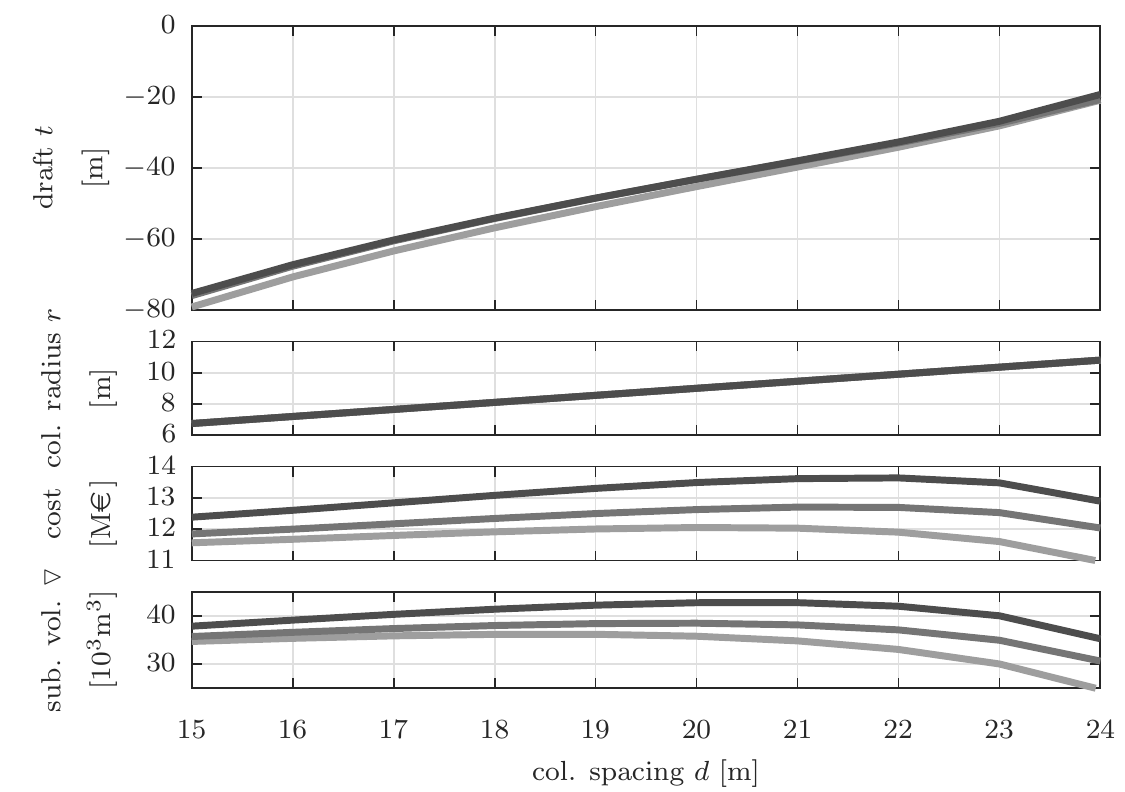}}
\put(0.129,0.38){
\includegraphics[scale=0.15]{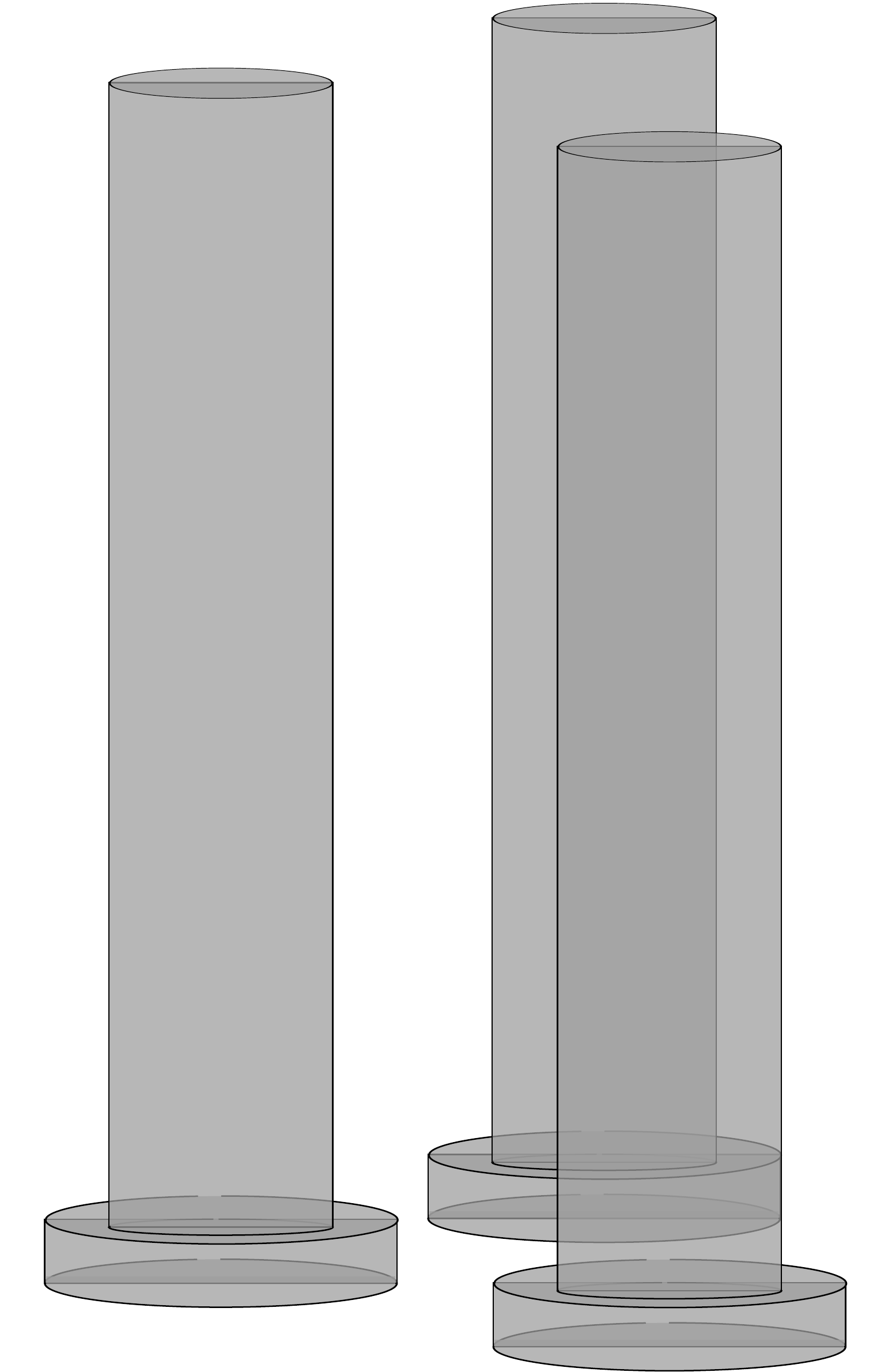}}
\put(0.37,0.47){
\includegraphics[scale=0.15]{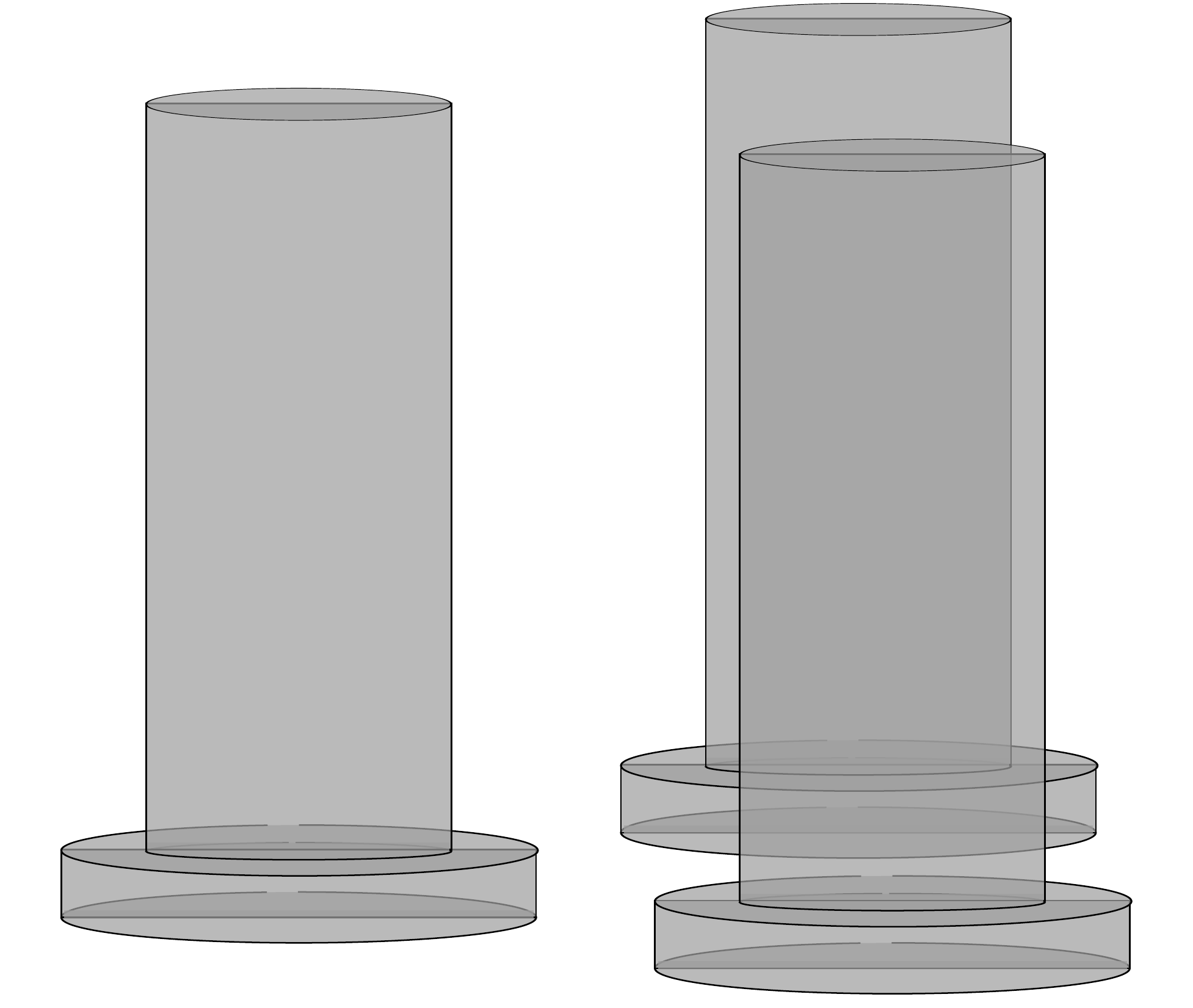}}
\put(0.69,0.58){
\includegraphics[scale=0.15]{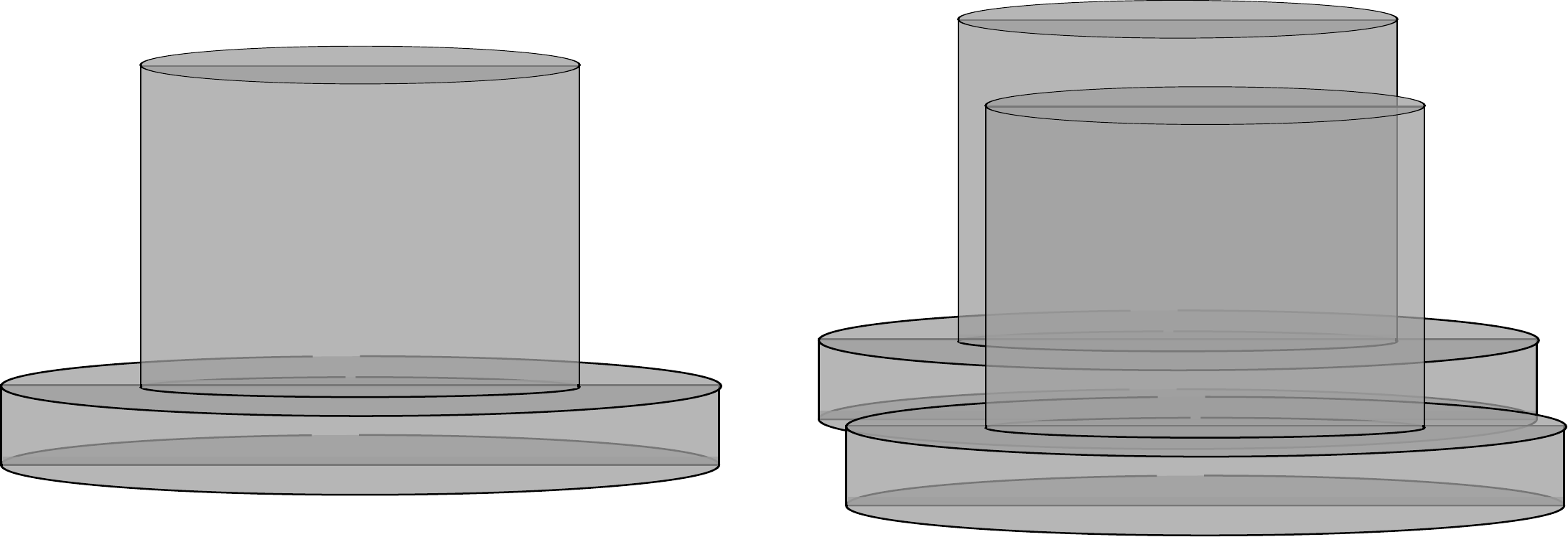}}
\end{picture}
\caption{Design space with two dimensions: column spacing from centerline and heave plate height. Heave plate height~$h_{hp}=[1.0,4.5,8.0]\,\si{m}$~(increasing darkness),~\cite{Lemmer2018a}.}\label{fig:DesignSpace}
\end{figure}

\section{Parametric Design Routines}\label{sec:parametricdesign}
This section addresses the design routines, which determine the dependent variables of Table~\ref{tab:parameters}. It starts with the structural design of the concrete floater with the steel tripod on top, followed by the mooring lines and the tower. Hydrostatic computations and the panel code calculations follow thereafter in Sections \ref{sec:designhst}--\ref{sec:designhydro}. The wind turbine controller is also parameterized and subject of Section~\ref{sec:designcontrol}. The wind turbine design, used for all platforms, is the \rwt of~\cite{Bak}. 

\subsection{Structural design}
Design routines, which were developed for the concrete columns, the heave plates and the steel tripod will be described in this section. The result of the structural design calculation is mainly the mass distribution, next to the shape of the hull, which will be input to the coupled simulation model. 
The assumed material properties, costs and wall thicknesses of the columns are listed in Table~\ref{tab:structure}, based on~\cite{Lemmer2016a}. 

\begin{table}[htp]
\centering
\caption{Structural design assumptions,~\cite{Lemmer2018a}.}\label{tab:structure}
\vspace{0.5em}
\begin{tabular}{lcr}
\hline
Parameter & Unit& Value\\
\hline
Reinforced concrete average density & [\si{kg/m^3}]& 2750.0\\
Steel density & [\si{kg/m^3}]& 7750.0\\
Ballast density & [\si{kg/m^3}]& 2500.0\\
Processed steel cost & [\si{\EUR/t}]& 4500\\
Processed concrete cost & [\si{\EUR/t}]& 399\\
Concrete column wall thickness & [m]& 0.6\\
Heave plate upper and lower lid thickness & [m]& 0.4\\
\hline
\end{tabular}
\end{table}

\subsubsection{Steel tripod}
For the steel tripod, various \gls{fe} analyses were performed, covering the range of radial distances of the columns~$d$ as shown in Figure~\ref{fig:DesignSpace}. The results of these calculations allow a parameterization of the dimensions of the steel legs like the sheet metal thickness, the strut width and height. The calculations assume a static thrust force at the tower-top of~\SI{4605}{kN}, as given in~\cite[p.~61]{Bak}. The notch stress at the joint between the tower and the tripod struts is assumed to be the critical failure mode. The dimensions of the struts are determined such that the maximum notch stress is of comparable magnitude as the one resulting from the same calculation with a comparable commercial tripile. The method was first applied in the European project INNWIND.EU\footnote{\url{http://www.innwind.eu}, accessed on July 18, 2018.}, of which details can be found in~\cite{innwindd433}. The dimensions of the tripod of the minimum and maximum column spacing are shown in Table~\ref{tab:tripod}. 

\begin{table}[htb]
\centering
\caption{Parametric design of the TripleSpar steel tripod,~\cite{Lemmer2018a}.}\label{tab:tripod}
\vspace{0.5em}
\begin{tabular}{lcc}
\hline
& Min. column spacing & Max. column spacing\\
\hline\\[-0.7em]
&\includegraphics[width=0.2\textwidth]{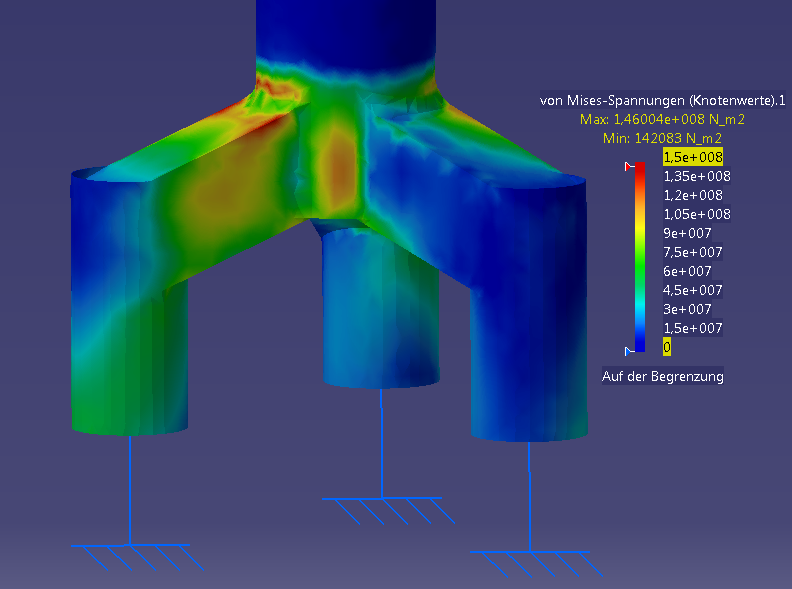} &\includegraphics[width=0.2\textwidth]{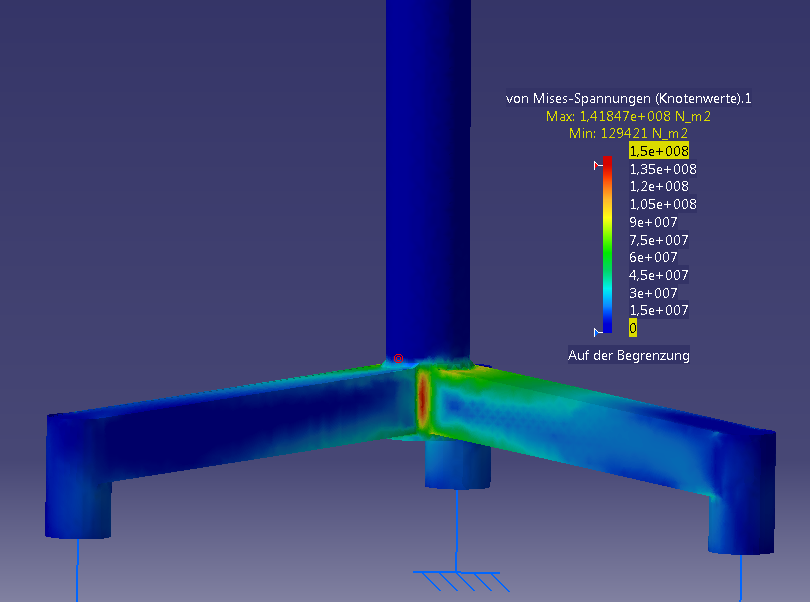} \\
&\def\svgwidth{0.2\textwidth}
\begingroup%
  \makeatletter%
  \providecommand\color[2][]{%
    \errmessage{(Inkscape) Color is used for the text in Inkscape, but the package 'color.sty' is not loaded}%
    \renewcommand\color[2][]{}%
  }%
  \providecommand\transparent[1]{%
    \errmessage{(Inkscape) Transparency is used (non-zero) for the text in Inkscape, but the package 'transparent.sty' is not loaded}%
    \renewcommand\transparent[1]{}%
  }%
  \providecommand\rotatebox[2]{#2}%
  \ifx\svgwidth\undefined%
    \setlength{\unitlength}{191.23984375bp}%
    \ifx\svgscale\undefined%
      \relax%
    \else%
      \setlength{\unitlength}{\unitlength * \real{\svgscale}}%
    \fi%
  \else%
    \setlength{\unitlength}{\svgwidth}%
  \fi%
  \global\let\svgwidth\undefined%
  \global\let\svgscale\undefined%
  \makeatother%
  \begin{picture}(1,0.6687753)%
    \put(0,0){\includegraphics[width=\unitlength]{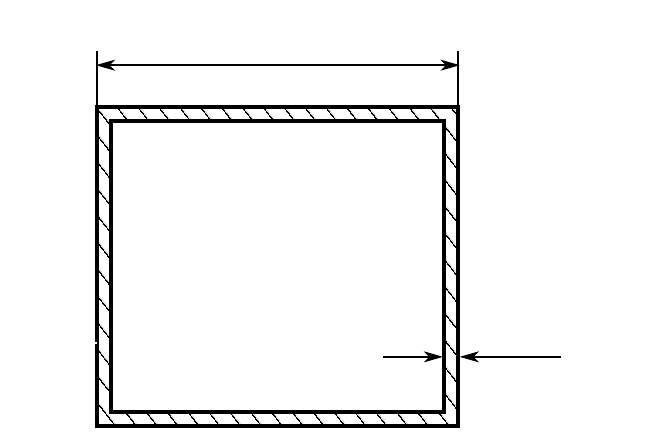}}%
    \put(0.71114888,0.15229567){\color[rgb]{0,0,0}\makebox(0,0)[lb]{\smash{\SI{50}{mm}}}}%
    \put(0.25996712,0.67467749){\color[rgb]{0,0,0}\makebox(0,0)[lt]{\begin{minipage}{0.31673018\unitlength}\centering \SI{5}{m}\end{minipage}}}%
  \end{picture}%
\endgroup%

&\def\svgwidth{0.25\textwidth}
\begingroup%
  \makeatletter%
  \providecommand\color[2][]{%
    \errmessage{(Inkscape) Color is used for the text in Inkscape, but the package 'color.sty' is not loaded}%
    \renewcommand\color[2][]{}%
  }%
  \providecommand\transparent[1]{%
    \errmessage{(Inkscape) Transparency is used (non-zero) for the text in Inkscape, but the package 'transparent.sty' is not loaded}%
    \renewcommand\transparent[1]{}%
  }%
  \providecommand\rotatebox[2]{#2}%
  \ifx\svgwidth\undefined%
    \setlength{\unitlength}{240.46384377bp}%
    \ifx\svgscale\undefined%
      \relax%
    \else%
      \setlength{\unitlength}{\unitlength * \real{\svgscale}}%
    \fi%
  \else%
    \setlength{\unitlength}{\svgwidth}%
  \fi%
  \global\let\svgwidth\undefined%
  \global\let\svgscale\undefined%
  \makeatother%
  \begin{picture}(1,0.66260165)%
    \put(0,0){\includegraphics[width=\unitlength]{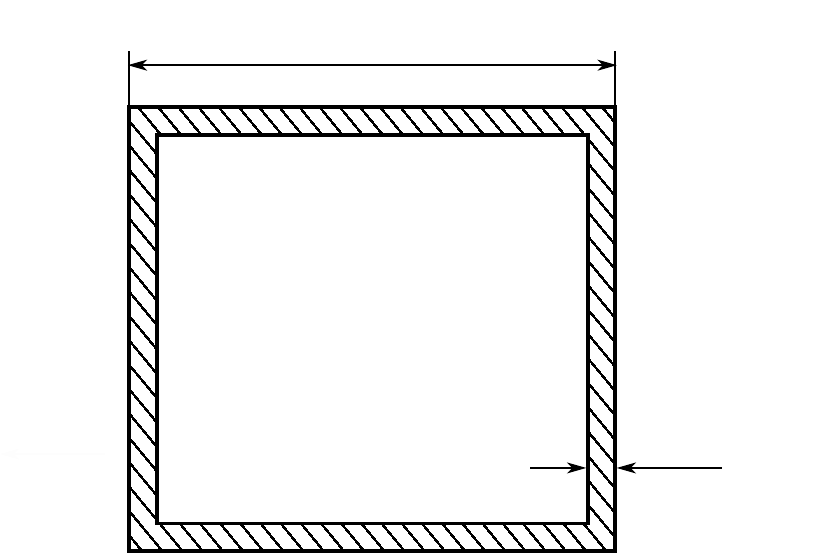}}%
    \put(0.77027796,0.11877151){\color[rgb]{0,0,0}\makebox(0,0)[lb]{\smash{\SI{60}{mm}}}}%
    \put(0.31164796,0.66729563){\color[rgb]{0,0,0}\makebox(0,0)[lt]{\begin{minipage}{0.25189412\unitlength}\centering \SI{7}{m}\end{minipage}}}%
  \end{picture}%
\endgroup%
\\
\hline
Column spacing 	& \multirow{2}{*}{\SI{10.0}{m}} & \multirow{2}{*}{\SI{35.0}{m}}	\\
(distance to center) &&\\
Strut width \& height	& \SI{5.0}{m} & \SI{7.0}{m}	\\
Steel wall thickness	& \SI{50}{mm} & \SI{60}{mm}	\\
Maximum stress	& \SI{146.0}{N/mm^2} & \SI{142.0}{N/mm^2}	\\
Tripod mass	& \SI{447}{t} & \SI{1716}{t}	\\
\hline
\end{tabular}
\end{table}

\subsubsection{Concrete columns and heave plates}
The concrete columns are of pre-stressed concrete following the example of the~AFOSP spar design, see~\cite{Molins2014a}. The cylindric walls have a constant thickness for all of the hull shape variations, see Table~\ref{tab:structure}. The heave plates are made out of reinforced concrete with the material properties given in Table~\ref{tab:structure}. The heave plates are concrete rings, flush mounted to the lower ends of the columns. This assumption is rather conservative and accounts for further compartmentation and reinforcements of the detailed design phase. 

\subsubsection{Mooring lines}
The properties of the mooring system can be found in~\cite{Lemmer2016a}. It was designed for the TripleSpar design~\cite{Lemmer2016a,Bredmose2017}, 
with two upwind lines and one downwind line with fairleads above~\gls{swl}, at~$z_\mathit{frlds}=\SI{8.7}{m}$ and a radial distance from the tower centerline of~$d_\mathit{frlds}=\SI{26}{m}$. This large distance makes it possible to keep the same mooring system for all column spacings~$d$ of Figure~\ref{fig:DesignSpace}.

\subsubsection{Tower}
The tower design is not varied in the optimization study but the parameters of the reference~TripleSpar are used~\cite{Lemmer2016a}. For \fowts it is important that the tower eigenfrequency does not coincide with the \gls{3p}-frequency range of the rotor. However, the considered platforms of Figure~\ref{fig:DesignSpace} show only little variations of the eigenfrequencies. This is mainly due to the hydrostatic design constraints (Section~\ref{sec:designhst}). Consequently, no adaptation of the tower design is considered in this work.  

\subsubsection{Cost estimation}
\fowt cost models have been subject of different research projects with a good overview and summary in~\cite{Benveniste2015,Lerch2018}. For this work, a cost of the processed material of the concrete heave plates and columns and the steel tripod, including both, material and manufacturing costs, is assumed. The values are in line with LIFES50+ Deliverable~4.3~\cite{Lemmer2016d} and can be found in Table~\ref{tab:structure}. The cost levels shown in Figure~\ref{fig:DesignSpace} are comparable to the ones of~\cite{Lerch2018}. Other costs than the ones mentioned, like installation, assembly and O\&M costs are not considered. It is noted that these values are rough approximations, which might change due to concrete shrinkage
and also due to existing price fluctuations over time.

\subsection{Hydrostatic design}\label{sec:designhst}
The hydrostatic restoring stiffness  in pitch-direction~$C_{55}$ is a driving property, which influences the~\fowt mean pitch angle under wind loads and therefore the power losses from misalignment, the maximum amplitudes and the eigenfrequencies. In other works, such as~\cite{Hall2013}, the value of~$C_{55}$ is free to vary in the design space and constraints on the pitch eigenfrequency are imposed. 
%

In this work, the hydrostatic restoring was constrained, following the perception that the hydrostatic restoring is crucial for the above-mentioned \fowt properties. A variation of~$C_{55}$ might result in small changes in cost and performance, while a large change of its value is unrealistic. The restoring $C_{55}=\SI{2.255e9}{Nm/rad}$ is set as constraint for all geometries of the design space. Depending on the free variables column spacing~$d$ and heave plate height~$h_{hp}$, the draft, as remaining variable, is determined such that this constraint is satisfied. A root-finding algorithm, connected to the structural design and the hydrostatic functions, solves this problem. The steady state pitch angle is about~$\beta_\mathit{p,rated}\approx\SI{3.0}{deg}$ at rated wind speed. It would change as a function of the fairlead vertical position, due to a different lever arm of the rotor thrust force. This vertical position, however, is constant throughout the design space.

\subsection{Hydrodynamic coefficients}\label{sec:designhydro}
A parameterized panel model has been set up to calculate the hydrodynamic coefficients with Ansys~Aqwa-Line. 

The column drag coefficient is kept constant for all designs and sea-states. It has been selected conservatively as~$C_D=0.4$. In the case of the heave plates, the drag coefficient is iterated based on the experimental data given in~\cite{Tao2008}. This data was parameterized as a function of~$\mathit{KC}$, which is possible because the Reynolds number and the surface roughness have little influence for shapes of sharp edges. For all designs, the vertical drag force is applied only at the lower (keel) surface of the heave plates and the bottom cross-sectional area of the heave plates is used for the calculation of the drag force. The iterative procedure is subject of the paper~\cite{Lemmer2018c}.
%

\subsection{Controller design}\label{sec:designcontrol}
As introduced in Section~\ref{sec:intro}, the blade pitch controller of the wind turbine for above-rated wind speeds is critical for the \fowt system stability. In order to ensure a properly tuned controller for each of the designs of Figure~\ref{fig:DesignSpace}, a method was developed to automate the design of a~\gls{siso} \gls{pi}-controller, depending on the linear dynamic model. This model will be introduced in Section~\ref{sec:slow}. The controller design procedure is subject of the companion paper~\cite{Lemmer2018d} and part of the thesis~\cite{Lemmer2018a}, which includes a comparison against a multivariable controller.

Figure~\ref{fig:FlowChart} of Section~\ref{sec:intro} shows on the right the iterative solution procedure to obtain the controller gains: It follows from the implementation of~\cite{Lemmer2018c}, which considers viscous drag coefficients of the heave plates that are a function of the $\mathit{KC}$-number, which is itself a function of the nodal response magnitude. The resulting platform viscous damping eventually determines how aggressive the wind turbine controller can be tuned. Thus, the viscous drag and the blade pitch controller gains are obtained iteratively.

The parameterized control design routine has proven to be valid throughout the entire design space of platforms, which was verified through simulations over all operating wind speeds. 
Especially the rotor speed and power overshoots are for all platforms within the design space and the metocean conditions of~\cite[Chap.~7]{Krieger2016} inside commonly used design constraints.

\section{Simulation Model}\label{sec:slow}
\begin{figure}[h]
\begin{minipage}[T][0.44\textheight][b]{1.0\textwidth}
\begin{minipage}[b]{0.35\textwidth}
\centering
\small
\def\svgwidth{1.0\linewidth}
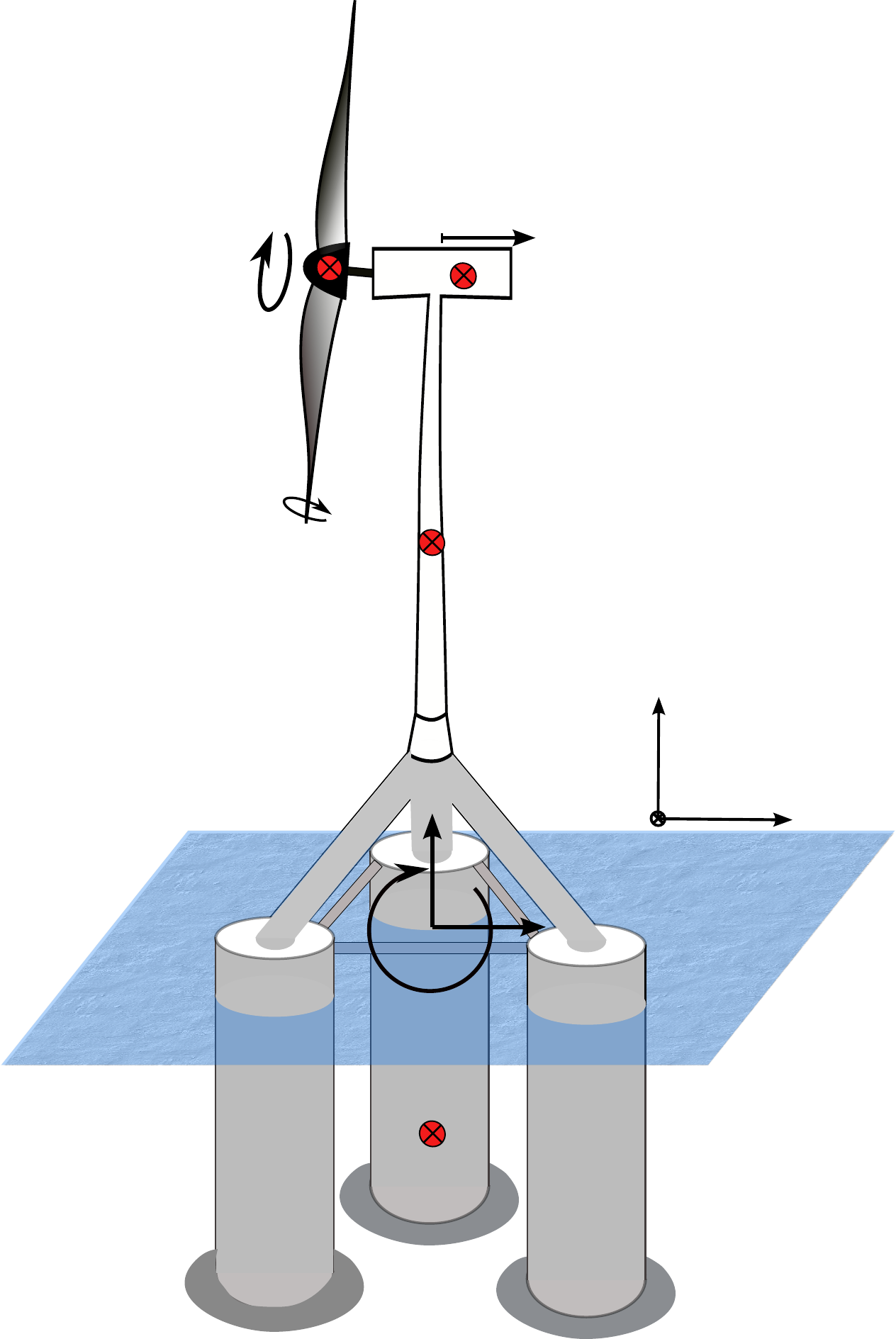%
\end{minipage}%
\hfill%
\begin{minipage}[b]{0.65\textwidth}
\renewcommand{\arraystretch}{0.5}
\small
\begin{itemize}
\item Structural dynamics: 6 \dofs, 2D motion
\item Aerodynamics: Quasi-static, rotationally sampled turbulence
\item Hydrodynamics: 1\textsuperscript{st}-order coefficients, Morison viscous drag, parametric heave plate drag, simplified radiation, 2\textsuperscript{nd}-order slow-drift approximation
\item Mooring dynamics: Quasi-static
\item Control: Model-based \gls{pi}
\item Time-domain \& frequency-domain
\end{itemize}
\vspace{0.5em}
\end{minipage}%
\captionof{figure}{Low-order simulation model \gls{slow} properties. Reprinted from~\cite{Lemmer2018c} with permission from MDPI, 2018.}\label{fig:TripleSparSketch}%
\end{minipage}
\end{figure}

The numerical \fowt model used for the computation of the load response in this work was described in~\cite{Lemmer2018a,Lemmer2018e}. Figure~\ref{fig:TripleSparSketch} shows a mechanical sketch of the \slow model. The intended purpose of the model is to represent the overall rigid-body and elastic dynamics of the system. It shall, on the other side, disregard all effects which increase complexity without contributing to the main system response. Through the improvement of computational speed, an efficient calculation of the load response for all designs is possible.
The model derivation starts with physical (as opposed to black-box) models, keeping the inherent nonlinearities. Subsequently, these models are linearized and the frequency-domain response spectra will be used here, which is subject of Section~\ref{sec:results}.

The structural model is derived using the theory of flexible \gls{mbs}. An advantage of the implemented algorithm is that the setup of the equations of motion is user-defined. Thus, the multibody structure can be adapted to the analyzed problem. In this work, the equations of motion have been set up for six \dofs in~2D: The floating platform rigid body in surge ($x_p$), heave ($z_p$) and pitch ($\beta_p$), a flexible tower with one generalized coordinate~$x_t$, the rotor speed~$\Omega$ and the blade pitch actuator~$\theta_1$ is selected, see Figure~\ref{fig:TripleSparSketch}.
These~\dofs are chosen with the aim of reducing the system to the most important forces and response dynamics. The largest environmental forces act usually in the direction of wind and waves~($x$). 

The aerodynamic model avoids an iterative solution of the lift and drag forces on the blade sections~\cite{Hansen2000}. Instead, the bulk flow across the entire rotor is considered and the rotor thrust and torque is calculated as function of the \gls{tsr} and the blade pitch angle, using the rotor thrust and power coefficient. This method neglects azimuth-dependent forcing, especially through wind shears. These are re-introduced through a rotational sampling of turbulence at the rotor rotational frequency. The method has proven to be a good tradeoff between accuracy and efficiency, especially for the optimization of the \fowt substructure, as shown in~\cite{Lemmer2018a}.

The hydrodynamic model uses the first-order panel code coefficients and a node-based Morison drag model for a reliable representation of the fore-aft viscous damping, as introduced in Section~\ref{sec:intro}. A significant improvement of the computational efficiency is achieved through the frequency-independent added mass and the neglected radiation damping. It is shown in~\cite{Lemmer2018a} that radiation damping is of little influence for the considered platforms, because the dominant forces at the frequencies of nonzero radiation damping are the first-order wave forces. Newman's approximation~\cite{Standing1986} is employed to account for low-frequency slow-drift forcing. The mooring forces result from a quasi-static model.

The nonlinear equations of motion are linearized about all operating points, resulting in a linear state-space description. The aerodynamics are linearized with a tangent approximation of the rotor coefficients through a central difference scheme. The quadratic drag is linearized through Borgman's method~\cite{Borgman1965a}, using the \gls{std} of the platform nodes, obtained from an iterative frequency-domain solution.
The linear model is only valid for small deviations of the states from the operating point. For this reason, the analyzed results focus on operational rather than extreme load cases. To prove the validity of the results, Section~\ref{sec:results} will include a comparison against the higher-fidelity FAST model.

\section{Load Calculation}\label{sec:results}
This section presents the results of load calculations in operational conditions across the design space. First, selected signal statistics will be compared, weighted with the probability of occurrence of each environmental condition over a lifetime of~20 years in the same way as~\cite{Smilden2017}. Thereafter, the frequency-domain response will be analyzed and the results will be verified through a comparison against the reference model FAST. A design indicator will be presented at the end of this section, which is able to predict the optimal response behavior. 

The environmental conditions of the project LIFES50+ are used with a significant wave height up to~$H_s=\SI{8.3}{m}$, see Table~\ref{tab:dlc}. For each wind speed, three different peak spectral periods~$T_p$ are considered, each with a probability of occurrence of 1/3 for the respective mean wind speed. One-hour simulations are performed with both time-domain models, cutting the transient for statistical evaluation.

\begin{table}[htb]
\caption{Environmental conditions for operational load cases of~\cite[Chap.~7]{Krieger2016}.}\label{tab:dlc}
\centering
\begin{tabular}{lccccccc}
\hline\noalign{\smallskip}
\small
Wind speed~$\bar{v}$ $[\si{m/s}]$	& 5.0& 7.1& 10.3& 13.9& 17.9& 22.1& 25\\
Significant wave & \multirow{2}{*}{1.4}& \multirow{2}{*}{1.7}& \multirow{2}{*}{2.2}& \multirow{2}{*}{3.0}& \multirow{2}{*}{4.3}& \multirow{2}{*}{6.2}& \multirow{2}{*}{8.3}\\
height~$H_s$ $[\si{m}]$ &&&&&&&\\
Peak spectral & \multirow{2}{*}{5.0}& \multirow{2}{*}{5.0}& \multirow{2}{*}{5.0}& \multirow{2}{*}{7.0}& \multirow{2}{*}{7.5}& \multirow{2}{*}{10.0}& \multirow{2}{*}{10.0}\\
period~$T_{p1}$ $[\si{s}]$&&&&&&&\\
$T_{p2}$  $[\si{s}]$& 7.0& 8.0& 8.0& 9.5& 10.0& 12.5& 12.0\\
$T_{p3}$  $[\si{s}]$& 11.0& 11.0& 11.0& 12.0& 13.0& 15.0& 14.0\\
Probability $f$ [\%] &14.8  & 25.0  &  28.7  &  17.5 &   5.9  &  0.9 &   0.1\\
\noalign{\smallskip}\hline
\end{tabular}
\end{table}

\subsection{Operational load response}
The linear frequency-domain SLOW model of Section~\ref{sec:slow} is used to simulate the response to the metocean conditions of Table~\ref{tab:dlc} for thirty platform designs of the parameter space of Figure~\ref{fig:DesignSpace}. 
%
 
\begin{figure}[htb]
	 \centering
	\includegraphics[scale=1.0]{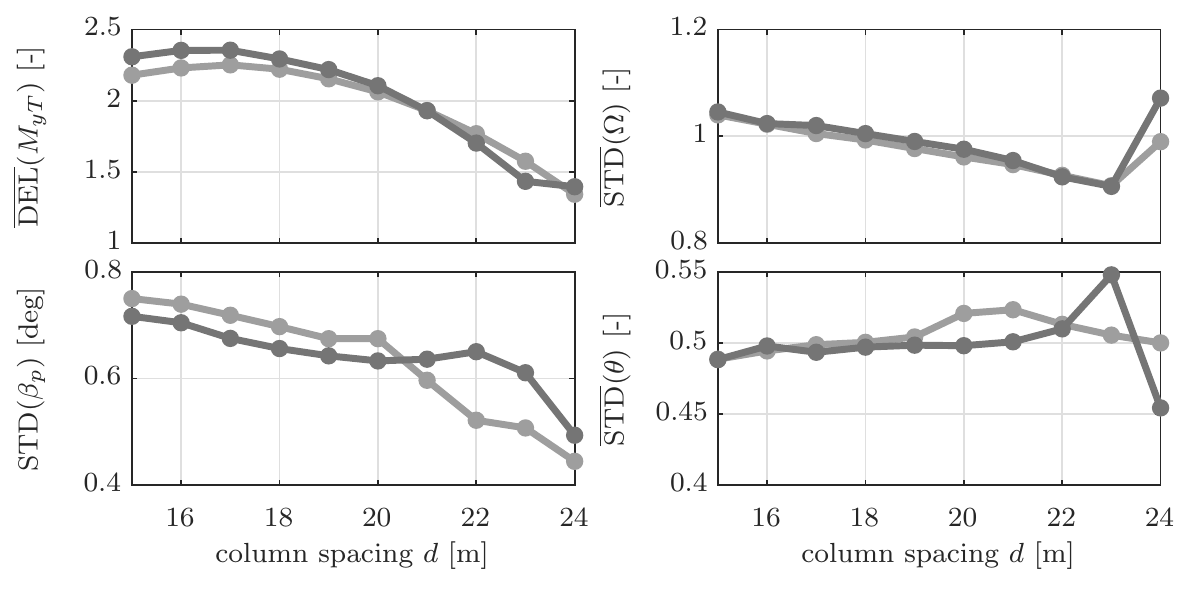}
	\caption{Linear response statistics of operational~\gls{dlc}1.2, of~\cite[Chap.~7]{Krieger2016}, weighted according to wind speed distribution. $\overline{\text{DEL}}$/$\overline{\text{STD}}$ are normalized results with corresponding~FAST onshore simulations of~\rwt with the same wind fields. Heave plate height~$h_{hp}=[1.0,8.0]\,\si{m}$~(increasing darkness),~\cite{Lemmer2018a}.}\label{fig:BruteForceWeighted}
\end{figure}

Figure~\ref{fig:BruteForceWeighted} shows the statistical results of the~IEC \gls{dlc} 1.2 \cite{IEC2007}, weighted with the Weibull probability density function of Table~\ref{tab:dlc}. The~\gls{del} of the tower-base fore-aft bending moment~$M_{yt}$ is calculated for a lifetime of~20\,years with a W\"{o}hler~exponent of~$m=4$. The~\glspl{std} are normalized with the corresponding values for the onshore \rwt with the same turbulent wind fields. For onshore turbines, the significant loading through the waves is not present and only the harmonic loads of the~\gls{1p} and~\gls{3p}-frequencies are the main fatigue drivers. It can be seen that the weighted~\gls{del} of~$M_{yt}$ has a minimum at the shallow-draft shape~($d=\SI{24}{m}$, $h_\mathit{hp}=\SI{4.5}{m}$). The same holds for the weighted~\gls{std} of the rotor speed~$\Omega$ (which is proportional to the power for above-rated winds) and the platform pitch angle~${\beta_p}$ (here the minimum is at the thinnest heave plate $h_\mathit{hp}=\SI{1.0}{m}$). The blade pitch activity~($\theta$) is rather constant over the design space. 

These results show a tower-base bending damage variation of more than~\SI{30}{\percent}. The fatigue damage is for the optimal design only \SI{50}{\percent} greater than for the onshore turbine, although the wave conditions are rather severe.  The estimated material cost of this design is not significantly higher than the one of the deeper draft, see Figure~\ref{fig:DesignSpace}, although the shallow-draft platform has a large column spacing of~$d=\SI{24}{m}$ and a column radius of~$r=\SI{10.8}{m}$ with heave plates of~$r_{hp}=\SI{17.4}{m}$.

\begin{figure}[htp]
	 \centering
	\includegraphics[scale=1.0]{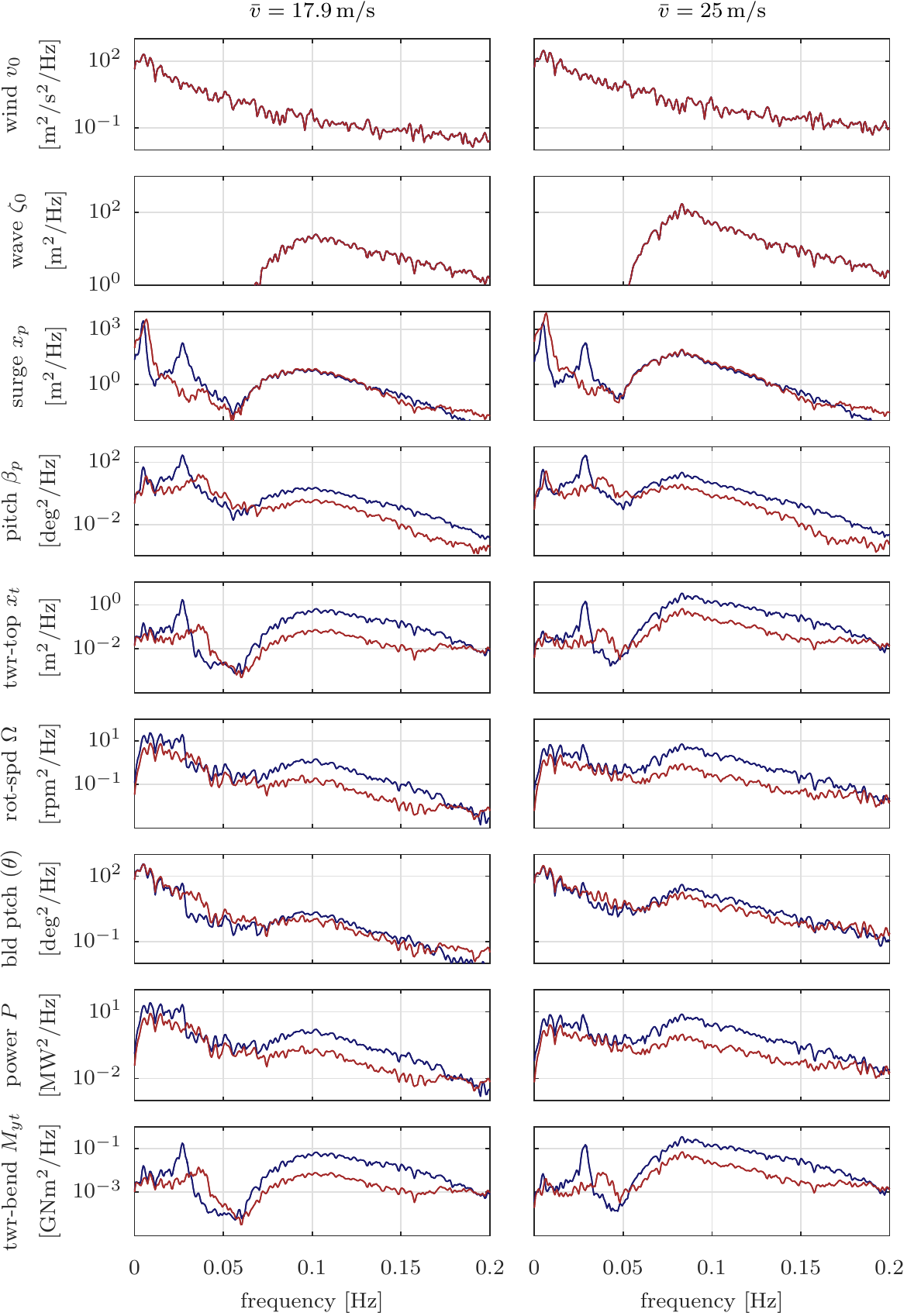}
	\caption{Comparison of deep-draft~(spacing $d=\SI{15}{m}$,~\clrI) and shallow-draft~(spacing $d=\SI{24}{m}$,~\clrII) platforms for~$\bar{v}_{hub}=\SI{17.9}{m/s}$~(left) and~$\bar{v}_{hub}=\SI{25.0}{m/s}$~(right) with~\gls{pi}-controller, computed by FAST,~\cite{Lemmer2018a}.}\label{fig:d15vsd24}
\end{figure}

Figure~\ref{fig:d15vsd24} shows a comparison of the response spectra of the deep-draft and the shallow-draft platforms. The shallow-draft shape has a larger difference-frequency and wind-induced response in surge at the lower end of the frequency axis. Except for this, the reason for the good performance, indicated by Figure~\ref{fig:BruteForceWeighted}, becomes clear: The shallow-draft platform has visibly smaller responses at the pitch eigenfrequency of~$f_\mathit{d,pitch}\approx\SI{0.04}{Hz}$ but, even more importantly, at the first-order wave frequencies of~$\SI{0.1}{Hz}$~($\bar{v}_{hub}=\SI{17.9}{m/s}$, left) and~$\SI{0.08}{Hz}$~($\bar{v}_{hub}=\SI{25}{m/s}$, right). This effect is noticeable for the platform motion response~$\beta_p$ but even more for the tower bending~($x_t$,~$M_{yt}$), the rotor speed~$\Omega$ and the electrical power~$P$. 

\subsection{Comparison of design model and reference model}
\begin{figure}[htb]
	 \centering
	\includegraphics[scale=1.0]{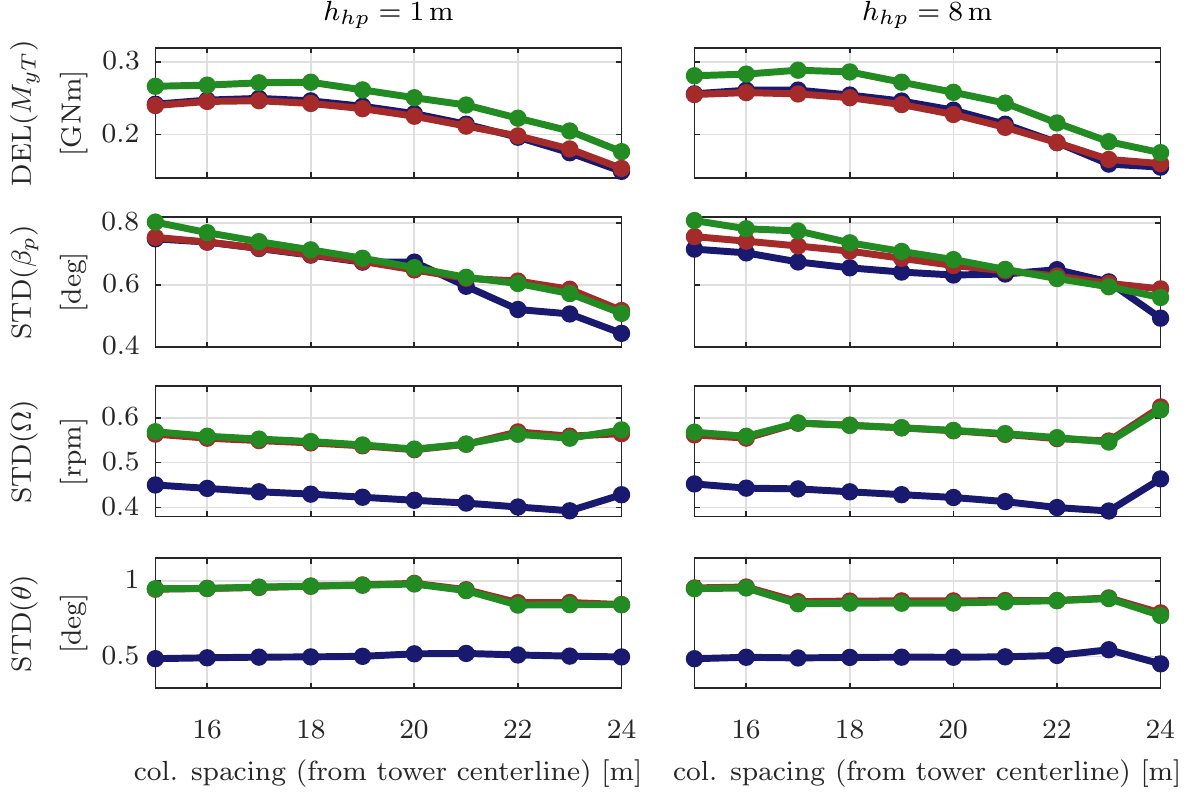}
	\caption{Response statistics of operational~\gls{dlc}1.2 with \gls{pi}-controller, weighted for all wind speeds with distribution. Linear SLOW model~(\clrI), nonlinear SLOW model~(\clrII), FAST~(\clrIII),~\cite{Lemmer2018a}.}\label{fig:BruteForceWeightedFast}
\end{figure}

Figure~\ref{fig:BruteForceWeightedFast} shows the same response signals as Figure~\ref{fig:BruteForceWeighted} with a comparison between the nonlinear and the linearized \slow models and the FAST model. The comparison shows the results of all column spacings~$d$ for the flattest heave plate~$h_{hp}=\SI{1.5}{m}$ and the thickest heave plate~$h_{hp}=\SI{8}{m}$. The qualitative optimum towards large~$d$~(shallow drafts) is predicted equally by all models. Consequently, the low-order model is generally suitable for the considered conditions and assessed response quantities. 

There is a constant concept-independent offset of the~\gls{del} between the \slow models and~FAST. The reason for this offset is the simplified actuator disk model. 
The excitation of the blades through wind shear is only re-introduced through a rotational sampling of the blade-effective wind at the nominal rotor speed, see~\cite{Lemmer2018e}. The offset of the rotor speed~$\Omega$ and blade pitch angle~$\theta$ is related to the linearized~\gls{slow} model. The linearized model underpredicts the variation of the rotor responses because of the nonlinear switching algorithm of the controller. The objective of the below-rated controller is to maximize power production, while the above-rated controller aims at limiting the power production. The switching between the control regions is a highly nonlinear effect, which cannot be modeled with the linear frequency-domain model. It is, however, modeled in the time-domain model, which agrees notably better with~FAST for~$\Omega$ and~$\theta$.
The platform pitch angle variation is well aligned among the time-domain models with a slight underprediction of~\gls{slow} for the platforms of deep draft. The linearized model underpredicts the pitch-\gls{std} for the shallow-draft platforms. This effect is distinct for the two heave plate heights~$h_{hp}$ and might therefore be due to the simplified radiation model, see~\cite{Lemmer2018e}. 

\subsection{Extreme load response}
\begin{figure}[htb]
	 \centering
	\includegraphics[scale=1.0]{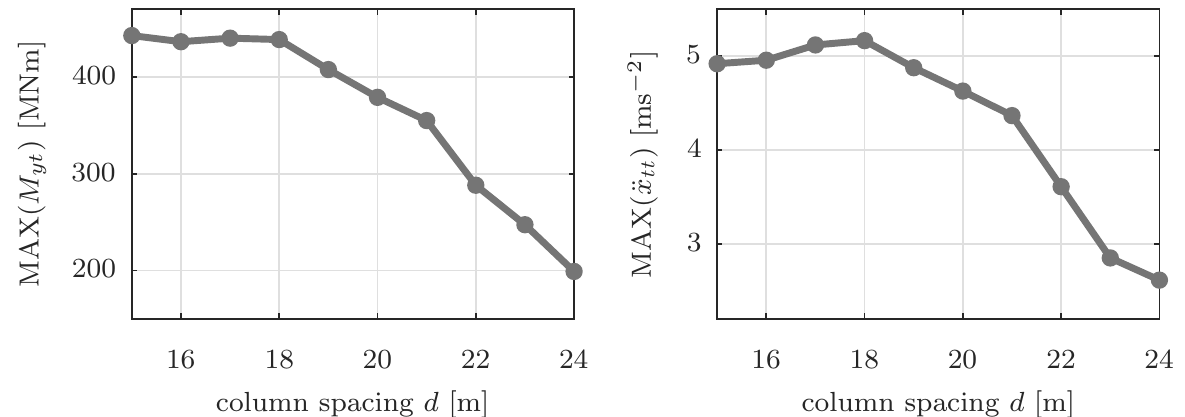}
	\caption{DLC 6.1 extreme sea state with 50-year extreme conditions of $\bar{v}_{50}=\SI{44}{m/s}$, $H_s=\SI{10.9}{m}$ and~$T_p=\SI{15}{s}$. Tower-base bending moment (left) and absolute tower-top acceleration (right) simulated with FAST. Heave plate height~$h_{hp}=4.5\,\si{m}$.}\label{fig:BruteForceDLC6d1} 
\end{figure}

The load response to IEC \gls{dlc} 6.1 with 50-year wind and wave conditions is presented in this section. The reference model FAST was used in this case because of large expected excursions, where especially the aerodynamic model of \slow might not predict reliable results, due to its simplifications. Turbulent wind and irregular wave timeseries were generated according to~\cite[p.~58, 64]{Krieger2016}. Figure~\ref{fig:BruteForceDLC6d1} shows the mean of the maxima of the tower-base bending moment~$M_\mathit{yt}$ and the tower-top acceleration~$\ddot{x}_\mathit{tt}$ (the absolute one, including platform accelerations) of three one-hour simulations. The results confirm the findings of Figure~\ref{fig:BruteForceWeighted}: The shallow-draft platform has small tower-top accelerations of less than $0.3g$. The LIFES50+ design basis~\cite{Ramachandran2017} would accept these values even during operation, while the limits for shut-down conditions are $\ddot{x}_\mathit{tt,max}=0.3g$.

\section{Results Analysis and Design Indicators}\label{sec:analysis}
A key to understanding the underlying effect for this remarkable ability of the shallow-draft platform~($d=\SI{24}{m}$) to reject disturbances is the analysis how the system responds to sinusoidal disturbances, meaning regular waves of different frequencies or harmonically oscillating wind speed. The ability to reject these disturbances will be quantified with a new design indicator at the end of this section.

\subsection{Harmonic response to wind and waves}
Figure~\ref{fig:HarmonicResponse} shows the harmonic response to wind and wave forces. The lines, representing each a different frequency, stand for the~\fowt centerline response. The horizontal distance of each point on the lines to the $y$-axis indicates the amplitude of oscillation of the system response at all elevations~$z$ along the centerline to sinusoidal wind~(upper) and wave excitations~(lower). The analysis is made with the low-order linearized model of Section~\ref{sec:slow} with first-order hydrodynamics, quasi-static mooring lines, an elastic tower, aerodynamics and the tuned~\gls{pi}-controller at~$\bar{v}_{hub}=\SI{13.9}{m/s}$. The centerline amplitude~$|\Phi_{x,i}(z,\omega)|$ results from the disturbance transfer function or \gls{rao}~$G_{di,x_k}(\omega)$ from wind ($i=1$) and waves ($i=2$) to the states surge, pitch and tower-bending, represented by~$x_k$ as
\begin{equation}
|\Phi_{x,i}(z,\omega)|=|G_\mathit{di,x_p}(\omega) + G_\mathit{di\beta_p}(\omega)\,z + G_\mathit{di,x_t}(\omega)\varphi_x(z)|.
\end{equation}
The tower shape function is denoted by~$\varphi_x(z)$.

\setlength{\unitlength}{\textwidth}%
\begin{figure}[htp]
\begin{picture}(0.9, 1.35)
\put(0,0){
\includegraphics[scale=1]{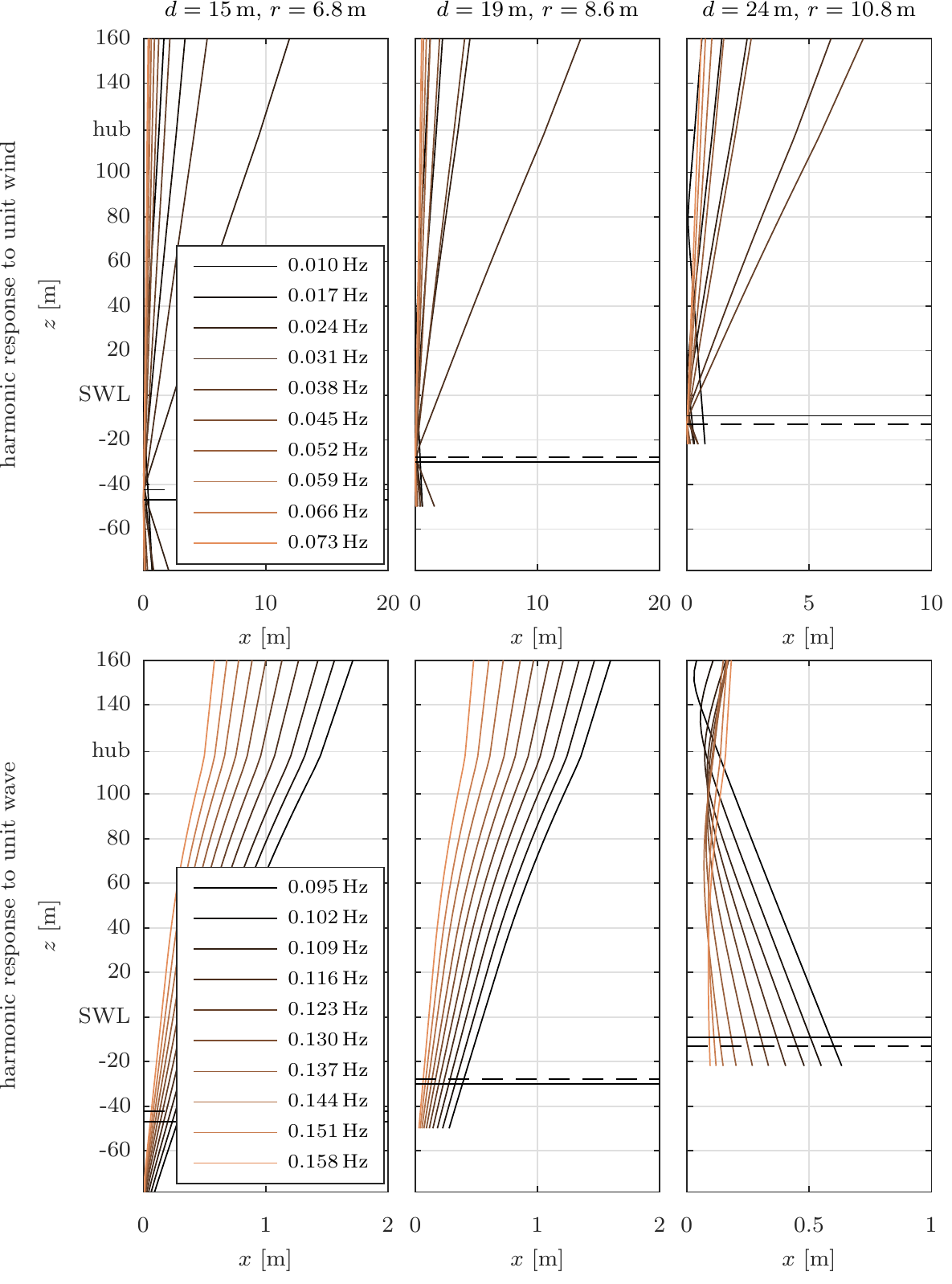}}
\put(0.25,1.1){
\includegraphics[scale=0.1]{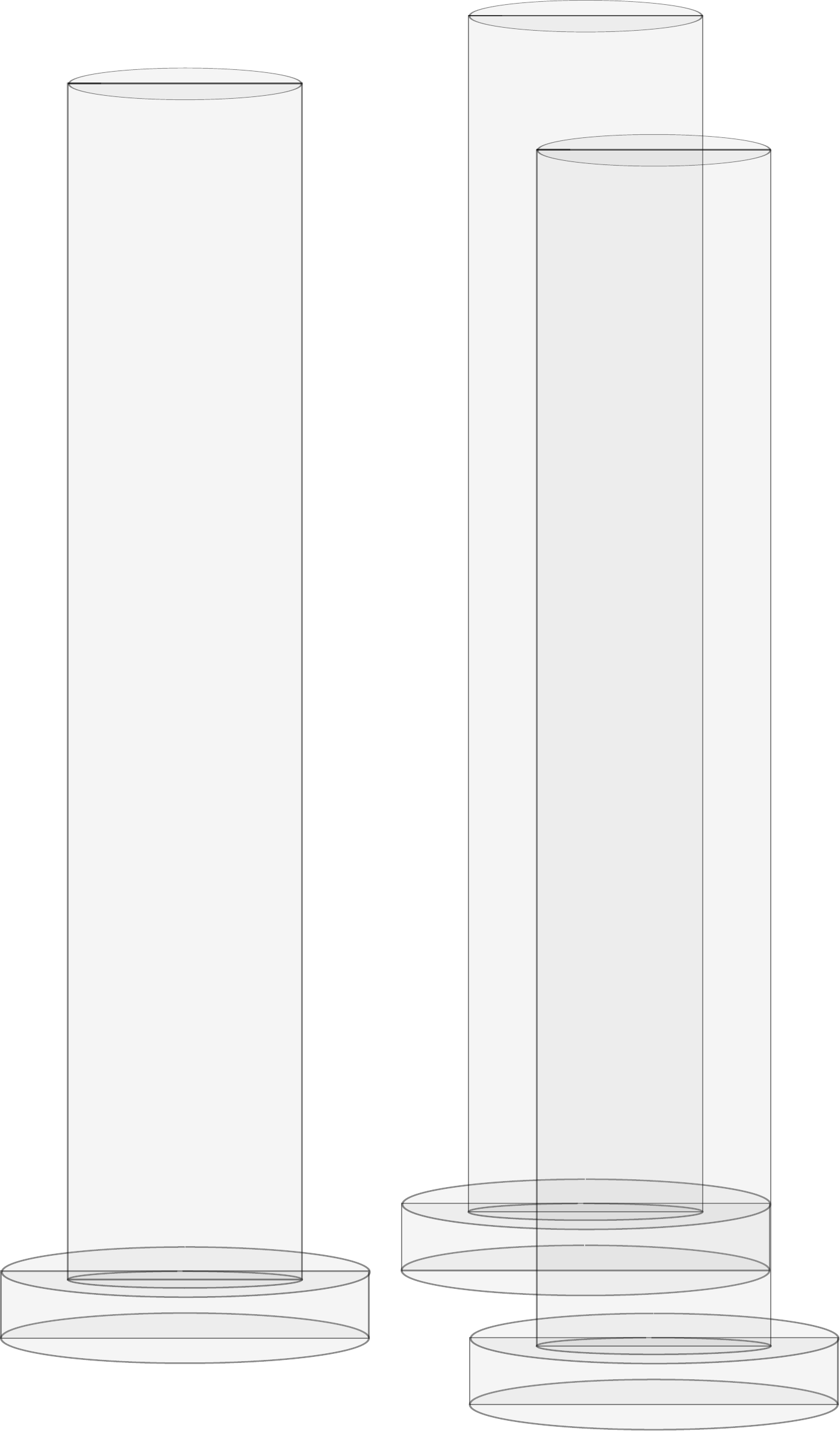}}
\put(0.52,1.15){
\includegraphics[scale=0.1]{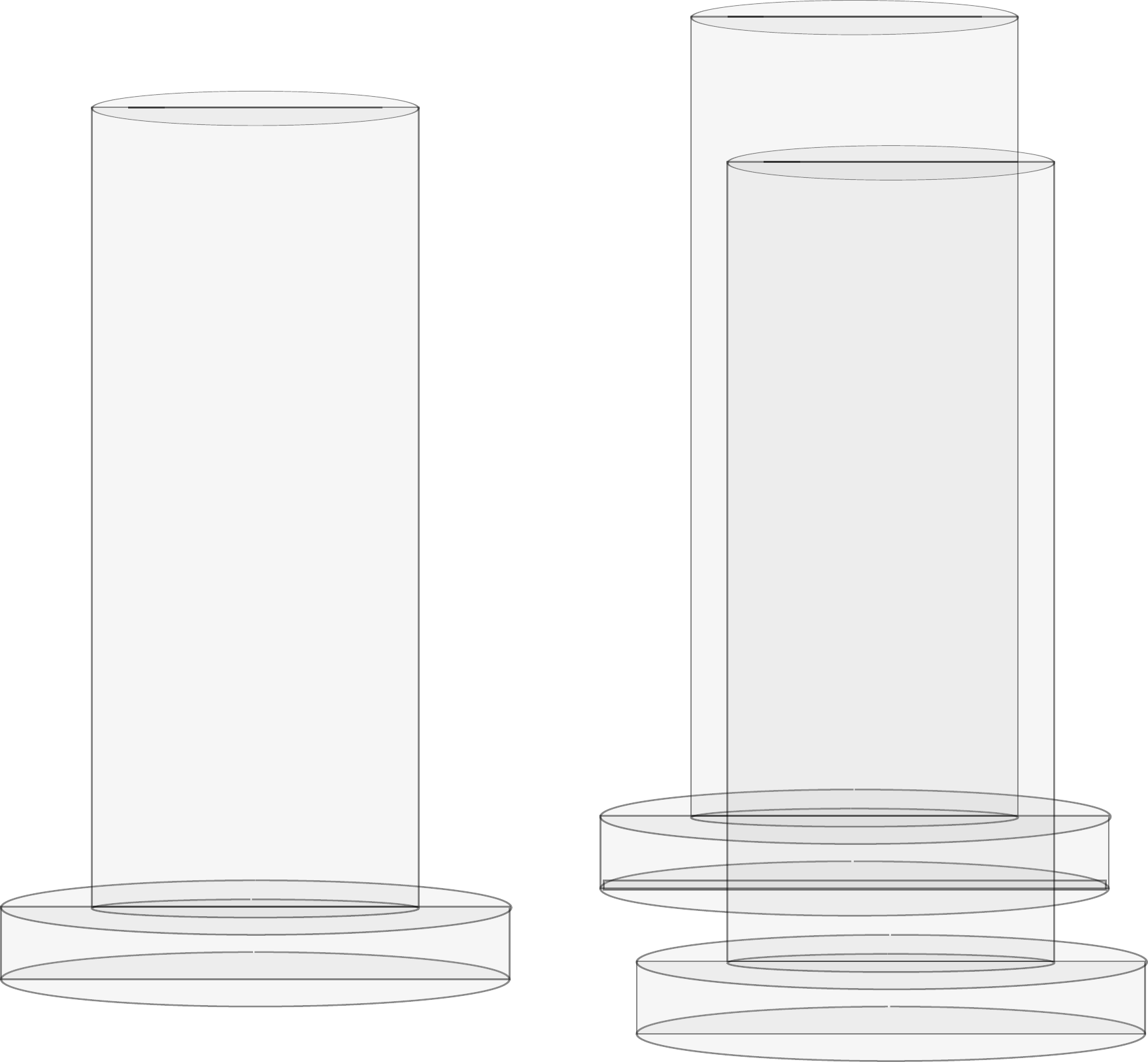}}
\put(0.78,1.2){
\includegraphics[scale=0.1]{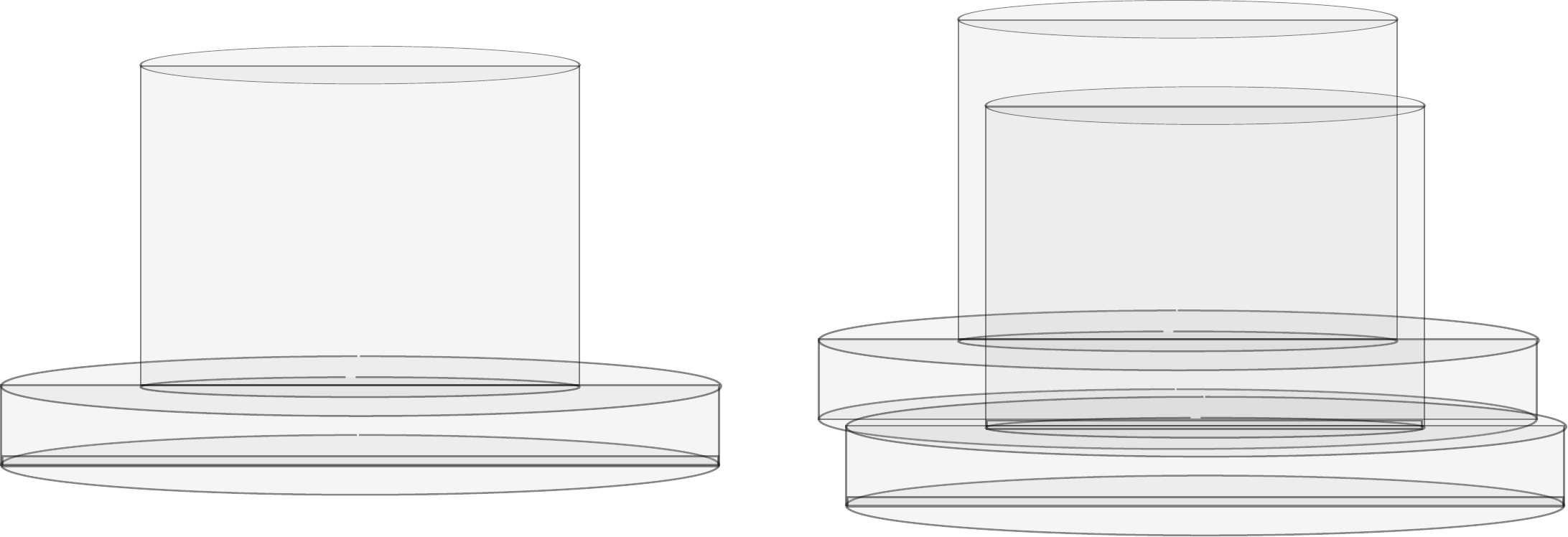}}
\end{picture}
	\caption{Harmonic response of~\gls{fowt} system over typical wind frequencies~(top) and wave frequencies~(bottom) at~$v_0=\SI{13.9}{m/s}$ with heave plate height~$h_{hp}=\SI{4.5}{m}$. Solid black line: \gls{fowt}~overall center of mass, dashed black line: center of buoyancy,~\cite{Lemmer2018a}.}\label{fig:HarmonicResponse}
\end{figure}

The solid black horizontal lines indicate the overall~\gls{cm} of the~\fowt and the dashed line indicates the center of buoyancy. For ships, the instantaneous center of roll motions is usually the metacenter,~\cite[p.~62]{Fossen2011}. Here, the instantaneous center of rotation in pitch is the vertical location on the platform with near-zero amplitude. The response to harmonic wind excitations shows a center of rotation equal to the~\gls{cm}. The same behavior is generally observable for all platforms.

The lower part of Figure~\ref{fig:HarmonicResponse} shows the harmonic response to waves. Here, the instantaneous center of rotation is below the platform-\gls{cm} for the platforms of deep draft. The platform of the lowest draft~($d=\SI{24}{m}$) has a different behavior: Its instantaneous center of rotation is at higher elevations, near the rotor hub. This is remarkable because it means that the hub does not respond with a horizontal movement to wave excitations. Consequently, the wind turbine power production is almost not affected by the waves and the rotor does not see any platform-induced relative wind speed. The shallow-draft platform responds in a way that surge is positive, when pitch is negative. Thus, these~\dofs are out-of-phase, as opposed to the other platforms, which have an in-phase response of surge and pitch to waves. Therefore, this optimality cannot be found when the surge or pitch~\gls{rao} is sought to be minimized. The same idea inspired the developers of a \gls{tlp} platform to arrange the taut mooring lines, such that the system is forced to rotate about the hub, see~\cite{Melis2016}. The fact that this behavior can be obtained for semi-submersibles has not been published yet. 

The cause leading to this response behavior are the Froude-Krylov pressures, integrated over the hull surface. The coupled magnitude and phases of the force-\gls{rao} vector~$\vek{X}(\omega)$ are for the optimal shape such that the presented response of the \fowt system is obtained. Semi-submersibles are predestined for a ``shaping'' of the force-\gls{rao} because the floater members see different phases of the waves and, additionally, the horizontal (heave plates) and vertical (columns) surfaces experience different phases of the Froude-Krylov pressures, which allow a counter-phase surge and pitch forcing. 

Eventually, the shallow-draft platform motion is not constrained but a motion response to mitigate the wave pressures is allowed. Important is the fact that the allowed motion response is favorable, such that the transmissibility of the wave forcing towards the wind turbine is reduced. The smaller tower-top motion yields smaller inertial forces of the \gls{rna}. The bending moments from the gravitational forces of the \gls{rna} are reduced because the platform pitch angle is also smaller for the shallow-draft design (Figure~\ref{fig:BruteForceWeighted}). The aerodynamic damping forces are not significant at the wave frequencies~\cite{Lemmer}. In conclusion, the stationary hub does not only yield a smooth power production but also reduced tower-base bending moments. Small adjustments in the optimal harmonic response shape can be expected when taking other outputs, like the mooring fatigue, into account.

The ``counter-phase pitch response'' of Figure~\ref{fig:HarmonicResponse} is important for the controller design: 
Because of the minimal motion response at the hub, the power fluctuations are reduced. Also, the coupling of the substructure dynamics with the controller is reduced. Common controllers feed back the rotor speed deviation from the set point. If the rotor responds less to the wave forces, there will be less wave-induced rotor speed deviation and  the wave response will not be amplified by the controller as with regular \fowt platforms. It is emphasized that the harmonic response is not to be confused with the modal response to turbulence and slow-drift forces. This one can be well damped by wind turbine controllers, which feed back a fore-aft velocity signal, as in~\cite{Fleming2016}, \cite{Olondriz2017} or~\cite{Yu2018}. It was shown in~\cite{Fleming2016,Lemmer} that the system response to first-order waves cannot be damped by the controller because of their large magnitude. Consequently, the shaping of the hull which yields the favorable first-order wave response can be effectively complemented by a controller, which damps the low-frequency motion response.


The sensitivity of this behavior to the peak spectral wave period~$T_p$ is obvious from Figure~\ref{fig:HarmonicResponse}. Thus, the favorable design is site-dependent. It can be seen, however, that the sensitivity is not very pronounced, meaning that the instantaneous center of rotation does not move significantly as a function of the wave frequency. This shows a certain robustness of the response behavior in terms of the environmental conditions.

\subsection{Design indicator: Minimum required control action}
\begin{figure}[htb]
\centering
   \small
   \def\svgwidth{0.7\linewidth}
   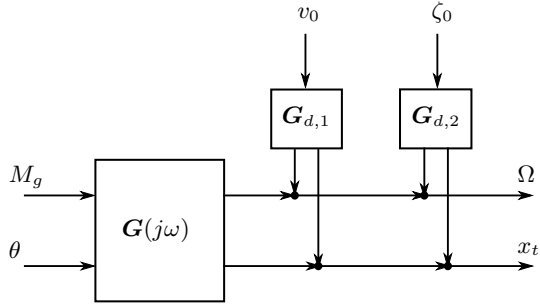
  	\caption{\gls{fowt} square plant with additive disturbances.}%
	\label{fig:gd}%
\end{figure}

The harmonic response of Figure~\ref{fig:HarmonicResponse} helps to understand the reason for the favorable behavior of the shallow-draft platforms. In this section, a new design indicator will be introduced to generally judge the goodness of a platform in terms of its induced wind turbine response. The idea is to quantify the necessary controller action to perfectly cancel the external forcing from wind and waves. The smaller the controller activity, the better the capability of the design to inherently reject disturbances itself. Figure~\ref{fig:gd} shows the linear~\fowt system transfer function in the Laplace domain~$\mat{G}(j\omega)$ with additive disturbance transfer functions~$\vek{G}_{d,i}$ for the control inputs generator torque and blade pitch angle~$\vek{u}=[M_g, \theta]^T$ and the outputs rotor speed and tower-top bending~$\vek{y}=[\Omega, x_t]^T$. Both, the plant~$\mat{G}(j\omega)$ and the disturbance transfer function for wind~$G_{d,v_0}(j\omega)$ and the one for waves~$G_{d,\zeta}(j\omega)$, collected in~$\vek{G}_d$, are obtained from the linear \slow model and scaled as in~\cite{Lemmer}. This is convenient as it means that a response~$y_i\geq1$ exceeds the design limits, see~\cite{Skogestad2005}.

\begin{figure}[htp]
	 \centering
	\includegraphics[scale=1.0]{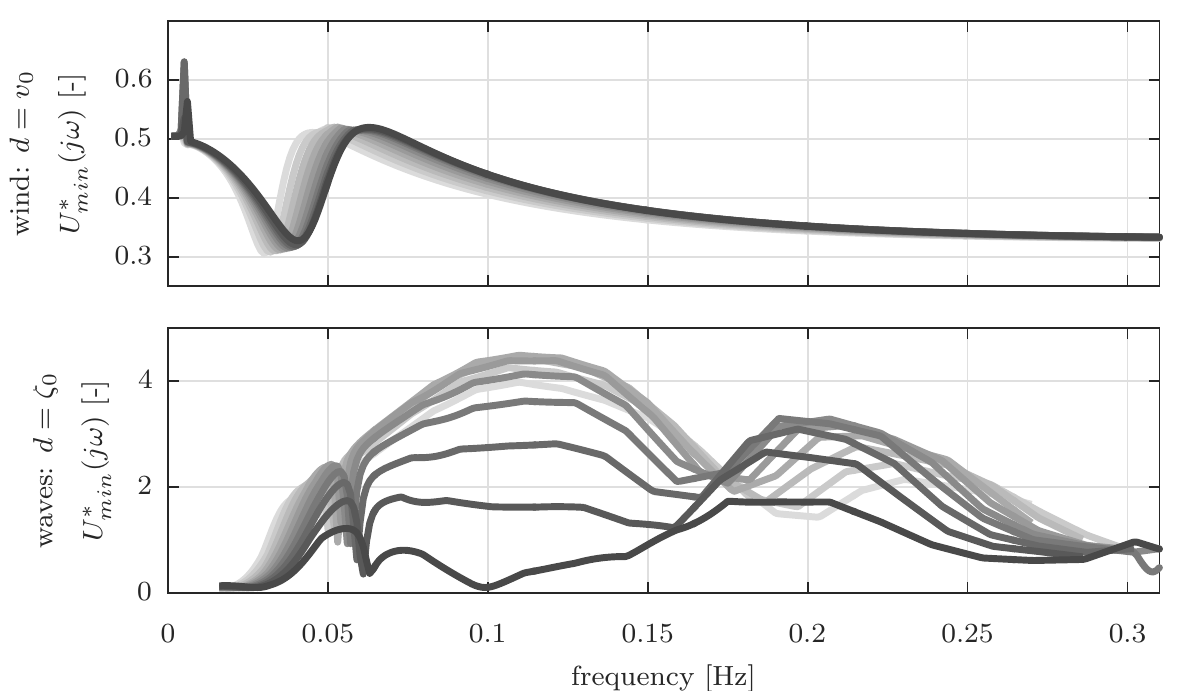}
	\caption{Minimum required control input magnitude~$U_{min}^*$ to perfectly reject wind~(top) and wave~(bottom) disturbances assuming the strongest combination of control inputs~$\vek{u}=[M_g,\:\theta]$ for platforms of different column spacings~$d=[15(1)24]\,\si{m}$~(increasing darkness) and~$h_{hp}=\SI{4.5}{m}$,~\cite{Lemmer2018a}.}\label{fig:Umin_designspace}
\end{figure}

In order to quantify the impact of disturbances on a linear system, Skogestad~\cite{Skogestad1996} calculates the 2-norm of the control input, theoretically necessary to perfectly cancel the disturbances. This analysis does not consider the controller dynamics but only the actuator authority on the system response. The required control input~$\vek{u}$ can be obtained, given a disturbance~$\vek{d}$, through system inversion as
\begin{equation}
\vek{u}=-\mat{G}^{-1}(j\omega)\vek{G}_d(j\omega)\vek{d}.
\end{equation}

The least necessary actuation\footnote{Calculated for both control inputs~$\vek{u}$, assuming the most effective combination of them, through a \gls{svd} of~$\mat{G}$. $U_{min}^*(\omega)$ is the 2-norm of the control input vector.} has been calculated for the different platforms for the wind excitation~$d=v_0=1$ and wave excitation~$d=\zeta_0=1$ independently in Figure~\ref{fig:Umin_designspace}. The part from wind excitation~(upper part of Figure~\ref{fig:Umin_designspace}) does not show a large variation over the different platforms, except for the changing eigenfrequencies. It is different for the wave excitation: The best performance (equal to the least required actuation~$U_{min}^*$) results here for the largest column spacing~$d=\SI{24}{m}$. 
Consequently, the assessment of the minimum required control input can well predict the optimality of the semi-submersibles, seen in the Figure~\ref{fig:BruteForceWeighted} and~\ref{fig:HarmonicResponse}. 

This indicator allows a general quantification of the ability of a \fowt platform to support a wind turbine without transmitting detrimental external excitation forces. It is the basis for a platform design, particularly adapted to the requirements of the wind turbine.

\paragraph{Parameters of optimum design} 
A FAST model of the optimum shape can be downloaded~\cite{Lemmer2018b}. The hull shape has a column radius of $r=\SI{10.81}{m}$, a spacing from the centerline of $d=\SI{24}{m}$ a heave plate thickness of $h_{hp}=\SI{4.5}{m}$ a draft of $t=\SI{21.94}{m}$. The platform has a mass including ballast of $m=\SI{3.115e4}{t}$ and a center of mass below \gls{swl} of~$z_\mathit{cm,ptfm}=\SI{13.36}{m}$.

\section{Conclusions}
\noindent
The study can be summarized by two parts: 

First, an integrated optimization procedure for semi-submersible \fowts was set up. A small design space of a three-column semi-submersible with heave plates was defined, ranging from a slender deep-drafted geometry to a shallow-draft one with large column diameter. A tailored coupled simulation model of low order was employed to calculate the system response, but also for the design of a wind turbine controller that is tuned for each platform design. As a result of the optimization, the design of lower draft gave a remarkable improvement of the response of the tower-base bending.

Second, a detailed analysis of the results revealed a particularly favorable response behavior to first-order waves, which can be represented by a new design indicator. The identified optimum is mainly due to an efficient design of the hull shape, yielding a favorable response behavior: The harmonic response function, derived from a linearized low-order model, revealed the dynamic characteristics of the optimum shape. It shows almost no fore-aft motion at the rotor hub, which means that the entire system, subject to wave loads, rotates about this point. As a result, the hub is almost stationary and the fluctuations of the power, the rotor speed, generator torque and blade pitch angle can be significantly reduced. An interpretation of this is that the surge and pitch response are out-of-phase, yielding a positive surge displacement when the pitch angle is negative. 

In order to consider this optimal dynamic behavior of semi-submersibles in future design practices, a new performance indicator was developed: The ``least required actuator action'' is considered, the action of the blade pitch angle and generator torque. Their magnitude, necessary to reject the wave loads, is an indicator, which can effectively predict how well a floating platform is able to inherently reject disturbances itself, without transmitting them to the wind turbine. 

The study showed that it is possible to design a \fowt platform, specifically suitable to carry a wind turbine, minimizing the effect of the waves on the power production. While the first-order wave loads cannot be mitigated by the controller, a further improvement of the low-frequency response behavior through the controller is possible. The presented harmonic response graph can be a good means for the selection of additional feedback loops. 
Although the cost model of this work is simple, the work follows the idea that a smooth behavior in harsh met-ocean conditions is a key for a sustainable and low-cost \fowt design.

Further studies will address the impact of the counter-phase response on the mooring line loads and integrate them into an optimization process. 
%

\section*{Acknowledgements}
The research leading to these results has received funding from the European Union's Horizon 2020 research and innovation programme under grant agreement No. 640741 (LIFES50+).
\bibliography{./library}

\end{document}